\documentclass[letterpaper,11pt]{JHEP3}
\usepackage{amsmath}
\usepackage{cite, amssymb}
\usepackage{epsfig}
\def\be{\begin{equation}}
\def\ee{\end{equation}}

\def\half{\frac{1}{2}}

\title{Holographic quantum liquids in 1+1 dimensions}
\author{Ling-Yan Hung and Aninda Sinha \\ {\it Perimeter Institute for Theoretical Physics, Waterloo, Ontario N2L
2Y5, Canada} \\ \vskip .5cm {\rm E-mail:}\ \ {\tt
jhung,$\,$asinha@perimeterinstitute.ca}}

\abstract{In this paper we initiate the study of holographic quantum liquids in 1+1 dimensions. Since the Landau
Fermi liquid theory breaks down in 1+1 dimensions, it is of interest to see what holographic methods have to say
about similar models. For theories with a gapless branch, the Luttinger conjecture states that there is an effective description of the physics in terms of a Luttinger liquid which is specified by two parameters. The theory we consider is the defect CFT arising due to a probe D3 brane in the AdS
Schwarzschild planar black hole background. We turn on a fundamental string density on the worldvolume. Unlike
higher dimensional defects, a persistent dissipationless zero sound mode is found. The thermodynamic aspects of
these models are considered carefully and certain subtleties with boundary terms are explained which are unique
to 1+1 dimensions.
Spectral functions of bosonic and fermionic fluctuations are also considered and quasinormal
modes are analysed. A prescription is given to compute spectral functions when there is mixing due to the
worldvolume gauge field. We comment on the Luttinger conjecture in the light of our findings.}

 \preprint{}

\begin{document}
\tableofcontents
\section{Introduction}

AdS/CFT \cite{Maldacena:1997re} is a powerful tool in extracting information about the strongly coupled limit of a conformal field
theory. The correspondence has been extended to the finite temperature limit\cite{Witten:1998zw}. In particular, there is recent
interest in understanding the thermodynamics, transport and spectral properties of strongly coupled low dimensional
systems, which are of interest in condensed matter physics \cite{Denef:2009yy,Herzog:2009md,Horowitz:2009ij,Gubser:2009gp,Gauntlett:2009dn,Faulkner:2009wj,Gubser:2009cg,Basu:2009qz,Alanen:2009cn} (and \cite{Starinetstalk,McGreevy:2009xe,Hartnoll:2009sz} for recent reviews).
Quantum liquids in one dimension have non-Fermi liquid properties and are thought to be of some relevance to the non-Fermi liquid properties of quasi-1D metals \cite{Voit} and high-T$_c$ superconductors \cite{anderson}.

Apart from its ubiquitous appearance and wide applications, one-dimensional Fermi liquids are of special
theoretical interests on their own right. In one dimension, Fermi liquids behave very differently from their
higher dimensional counterparts, which, in the weak coupling limit, can be very generally described by the
Landau Fermi liquid theory (see for example a textbook introduction in \cite{fermiliquidbook}). The Landau Fermi liquid theory asserts that the free fermion description of the
system is only mildly altered under the introduction of interaction, particularly if we are sufficiently close
to the Fermi-surface. Interaction between the fermions can be accounted for by introducing a renormalised mass
for the fermionic particles. These fermionic excitations decay in time, but sufficiently slowly as we get closer to
the Fermi-surface, so that they are reasonably well defined {\it quasi-particles} within their life-time. This
picture, however, completely breaks down in one dimension. The fundamental difference stems from the different
topologies of the Fermi-surfaces, which, in one-dimension, comprises of only two distinct points in momentum
space, instead of forming a continuous surface. This means that for an excitation from the ground state with a
sufficiently small momentum, the energy of the excitation is completely determined. In higher
dimensions a particle hole pair can acquire a continuous range of energies even for arbitrarily small total
momentum. In 1D, this means that for small momentum change about the Fermi-points the density fluctuations, which are
collective fluctuations, are eigenstates of the Hamiltonian. In fact, one can assume a quasi-particle
description in one dimension and calculate the particle decay rate, which diverges for arbitrarily small but
finite interactions, indicating a break-down of the quasi-particle picture.

\FIGURE[ht]{\begin{tabular}[h]{cc}
\includegraphics[width=0.2 \textwidth]{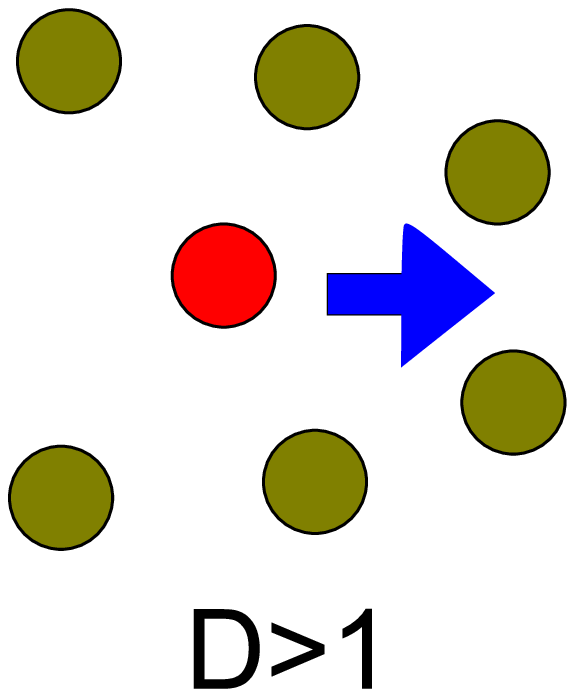} & ~~~~~~\includegraphics[width=0.4 \textwidth]{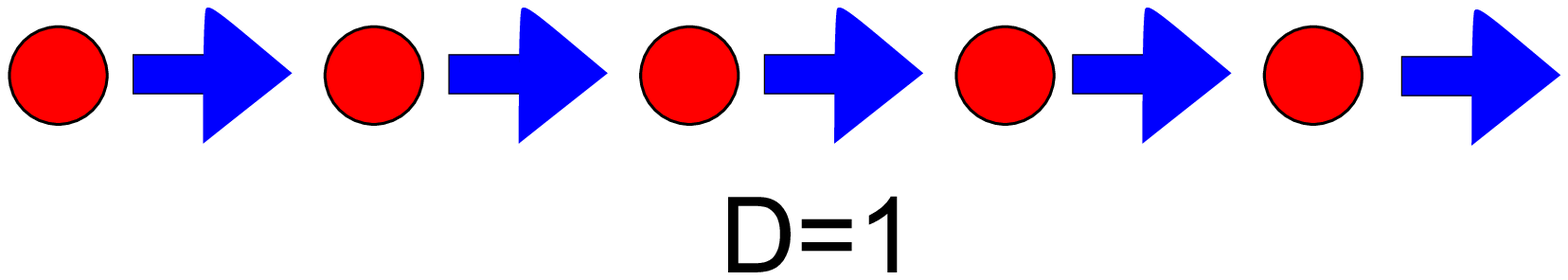}
\end{tabular}
\caption{In higher dimensions, Landau Fermi liquid theory advocates description using independent quasiparticles. In 1+1 dimension, collective excitations are crucial and the quasiparticle description breaks down.} \label{1d2d}}

The Luttinger liquid is
introduced to model the effects of interaction in one dimensional fermion system, in an analytically controlled
context\cite{Luttinger1}. For a spinless fermion, and ignoring back-scattering, which can be shown to be irrelevant in the low
energy limit, the model is reduced via bosonisation, to a free theory that can be exactly solved, which gives us
physical insight into the low energy spectrum\cite{Mattis,Luther,Haldane1}(and for a review see \cite{Voit,rao-2000}). The energy eigenstates, as anticipated above, are related to
density fluctuations of the fermion liquid by a Bogoliubov transformation. This is a massless excitation with
linear dispersion, which is one of the prominent features of the Luttinger liquid. In the more sophisticated
model of spin-half Luttinger liquid, the diagonalisation of the Hamiltonian leads to two independent
eigen-bosonic fluctuations, corresponding to the spin and charge densities fluctuations. They are again massless
excitations which often move with different speeds, depending on the coupling. This is the famous spin-charge
separation effect\cite{Haldane1,Voit,rao-2000}, which has recently been observed in experiments\cite{2005Sci30888A,2009Sci325597J}.

The {\it Luttinger conjecture} states \cite{Haldane1,Voit}  that any 1D model of correlated quantum particles (bosons or fermions) having a branch of gapless excitations will have as its stable low-energy fixed point the Luttinger model. The asymptotic low energy properties of the degree of freedom of this branch will be described by an effective renormalized Luttinger model characterized by only 2 parameters: a renormalized Fermi velocity $v$ and a renormalized stiffness constant $K$.
While successful in providing lots of physical intuition, the Luttinger liquid model involves lots of
simplification, without which analytic control would have been impossible. In fact lattice methods are used to extract effective parameters of the continuum model.  The holographic model is thus another
analytic handle towards understanding low-dimensional systems. The usefulness of these models is that the dynamic properties can be extracted more easily than lattice methods. For example, in higher dimensions the fermion sign problem makes it next to impossible to extract useful information about fermion correlation functions using the lattice method.

The model we will study in this paper is a $1+1$ dimensional defect CFT (dCFT). This is made from intersecting D3 branes and is the finite temperature generalization of \cite{erd}. The counterterms needed to compute correlation functions at zero temperature in this model were discussed in \cite{Karch:2005ms}. In order to be in a gapless phase, we will consider the phase described by the so-called black hole embeddings \cite{Mateos:2007vn}. Moreover, we will turn on a finite chemical potential so that the black hole embeddings are the only physical ones \cite{Kobayashi:2006sb}. Considering a defect CFT of this kind is also useful since it facilitates comparison with such theories in other dimensions. The crucial features of the D3-D7 model with a finite chemical potential which makes a 3+1 dimensional theory with fundamental matter are:
\begin{itemize}
\item A phase transition for low baryon densities and low temperatures \cite{Kobayashi:2006sb}.
\item In a certain momenta regime, quasiparticles become a good description for the excitations in the spectral functions \cite{Erdmenger:2008yj, Myers:2008cj, Mas:2008jz}. Spectral functions typically contain several Breit-Wigner resonances. A collective mode resembling the zero-sound was found \cite{Karch:2008fa, Kulaxizi:2008kv}. All modes have a dissipative part.
\item Dispersion relations for quasiparticles look like $\omega\sim a + b k^2$ for low $k$ and $\omega\sim v_c k$ for high $k$. At higher $k$, the peaks dissolve and the notion of quasiparticles is lost. $v_c$ is less than unity and a notion of a limiting velocity emerges \cite{Myers:2008cj, Ejaz:2007hg}.
\end{itemize}
These features are expected to hold even in the 2+1 dimensional dCFT \cite{Myers:2008me}.

The 1+1 dCFT at zero temperature and chemical potential is obtained by placing $N_f$ D3 branes in the background
geometry created by $N_c$ D3 branes
and considering the probe approximation $N_f \ll N_c$. The probe sees a $AdS_3\times S^1$ geometry. This
background is known to preserve half the original ${\cal N}=4$, $d=4$ supersymmetry and realizes a two
dimensional $(4,4)$ supersymmetry algebra. The conformal symmetry group is $SL(2,R)\times SL(2,R)$ inherited
from the $AdS_3$. The massless open string degrees of freedom correspond to a pair of ${\cal N}=4$ Yang-Mills
multiplet coupled to a bifundamental $(4,4)$ hypermultiplet living at the intersection. We send $N_c$ to
infinity keeping $N_f/N_c \ll 1$. This leaves a single ${\cal N}=4$ Yang-Mills multiplet coupled to the
bifundamental at a 1+1 dimensional defect. We will be concerned with the case where the probe sits at the origin
of the part of the AdS space transverse to its worldvolume. The modes corresponding to contracting the $S^1$ inside the $S^5$
saturates the Brietenlohner-Freedman bound for scalars in $AdS_3$ and hence are stable. The generating function
for the field theory is given by the classical action of superstring theory on $AdS_5 \times S^5$ coupled to a
Dirac-Born-Infeld theory on $AdS_3\times S^1$.

We will be considering this setup at finite temperature. Finite temperature is introduced by replacing the
$AdS_5$ by the 5-dimensional AdS-Schwarzschild black hole. This trick has been used to analyse the behaviour of
higher dimensional defect CFTs at finite temperature\cite{Kobayashi:2006sb,Myers:2007we,Mateos:2007vc,Mateos:2007vn,Myers:2008cj,Mas:2008jz,Erdmenger:2008yj,Shieh:2008nf,Kulaxizi:2008jx,Myers:2008me,Wapler:2009tr}. Part of this exercise is to see if we can probe the validity of the Luttinger conjecture using AdS/CFT. The other part is to probe differences between 1+1
dimensional theories and higher dimensional ones. To introduce a finite chemical potential and quark condensate, we look for a probe brane solution in the AdS black hole
background with a non-trivial brane profile and world-volume electric field.
The thermodynamics of the system is studied in detail.  In the case when the probe wraps the maximal circle (in the notation used
in this paper, $\chi=0$) we analytically derive the expressions for the
thermodynamic quantitites, where we find a heat capacity scaling as $T$ in the high temperature limit, as
expected in a $1+1$ d system, but $T^2$ in the low temperature region where the baryon density
becomes important. For the non-maximal case ($\chi\neq 0$) we turn
to numerics. The specific heat in this case, closely resembles that of the $\chi=0$ case.
We then compute holographically the Green's functions of various scalar, vector and fermionic operators by studying the world-volume fluctuations
of the probe brane about the background solution.

Our analysis is divided into several steps. To begin with, we study the fluctuations of the longitudinal electric field for
trivial embedding (where the brane passes straight through the horizon), very much in the spirit of
\cite{Karch:2008fa}. We find that there is a massless (zero-sound) mode with dispersion given by the conformal result
i.e. $\omega=\pm k/\sqrt{q} $, where $q$ is the defect dimension, equals to one in this case. We obtain this
result analytically in the hydrodynamic limit, and find no dissipation. We extend the study to larger
frequencies $\omega$ and momenta $k$ numerically, and find that the dispersion relation is unmodified, and
remain dissipationless. More surprisingly, while it is by now known that these massless (sound) modes in holographic defect models
disappear at finite temperatures for higher dimensional defects \cite{Kim:2008bv}, the massless mode in our 1D defect survives at finite temperatures, with identical dispersion as in the case of zero temperature.

We then generalise the investigation to the case of mixed excitations, where the fluctuations of the electric
field longitudinal to the defect are mixed with those of the embedding profile at finite momentum exchange, and
in the presence of background world-volume electric field and a non-trivial probe embedding. To deal with the
mixing, we explain in detail how the quasi-normal modes and spectral functions are obtained in principle\cite{Amado:2008ji}, and
implement the procedure numerically. It is interesting to see that the massless mode mentioned above survives
even in this limit without dissipation. There is also a mode corresponding to pure dissipation, which originates
from the profile fluctuation even before mixing is introduced. The dispersion of this mode is given
schematically by $\omega \sim -i (a + b k^2)$, for some constant $a$ and $b$ dependent on the background
electric field (controlling the chemical potential) and the probe embedding (controlling the quark mass and
condensate). The higher quasi-normal modes for bosonic excitations, however, behave roughly as the higher dimensional ones, with a
dispersion relation quadratic at small $\omega$ and $k$, and approaches that of the speed of light
asymptotically for large $\omega$ and $k$.

It is interesting to note that while our analytic expressions for the dispersions for the massless mode appear
to be independent of the charge density $d$, the limit $d\to 0$ at zero temperature is not
a smooth one. In fact when $d=0$ the fluctuations can be solved exactly analytically at zero temperature and we find
that the massless mode disappears. In fact the spectral function becomes a constant, for both the vector and
scalar modes. At zero $d$ and finite temperatures, the scalar fluctuations can again be solved exactly and the quasi-normal modes
form a discrete infinite tower of modes. The massless dissipationless mode that
appears in the longitudinal electric field fluctuations however,  remains intact here.

This paper is organized as follows. In section 2, we discuss the setup and specify our conventions. In section 3, we consider the thermodynamics of our embeddings and carefully analyse the counterterms needed for the calculations. In section 4, we turn to the analysis of zero sound. In sections 5 and 6 we consider the spectral functions for bosonic and fermionic fluctuations. We conclude with a discussion in section 6. Appendix A gives some analytic results for the thermodynamics. Appendix B has a proof that when the real part of the quasinormal mode is non-zero, then the imaginary part of the mode has to be necessarily positive indicating no instabilities. Appendix C has some explicit calculations for the Green functions for the mixed modes considered in section 5.

\section{D-brane configurations}\label{braneconfig}

We would like to begin with a review of the brane configuration, namely, the intersecting D3 systems, considered
in this note. Its zero temperature limit has been studied in depth in\cite{erd,Karch:2005ms}. We have $N_c$ D3
branes and $N_f$ D3 branes intersecting over a 1+1 dimensional domain with four relatively transverse
dimensions. The low energy effective field theory is given by a supersymmetric $U(N_c)\times U(N_f)$ Yang-Mills
theory on a 1+1 dimensional defect. In the large $N_c$ and large t'Hooft coupling $\lambda = 4\pi g_sN_c$ limit,
such that $N_f/N_c \ll 1$, one could replace the $N_c$ branes by the AdS geometry and treat the $N_f$ D3
branes as probes in the curved background. The excitations on the probe are then dual to the low energy
excitations on the 1+1 dimensional defect. Although the 1+1 dimensional defect will be the primary focus of this
paper, sometimes we will draw parallels to the well studied D3-D7, 3+1 dimensional setup as well. The
brane-scans for the two setups are
\begin{equation}
\begin{array}{ccccccccccc}
& 0 & 1 & 2 & 3 & 4& 5 & 6 & 7 & 8 & 9\\
\mbox{$~~$ $N_c$ D3:} & \times & \times & \times & \times & -& -& - & - & -&- \\
\mbox{1. $N_f$ D7:} & \times & \times & \times & \times & \times & \times
& \times & \times & - & -  \\
\mbox{2. $N_f$ D3:}& \times & \times &- & - & \times & \times
& - &  -&  -& -  \\
\end{array}
\label{D3D7}
\end{equation}

In the zero temperature limit both these theories are supersymmetric. We will consider the abelian case only and
leave the analysis for non-abelian effects for future work. At finite temperatures, the AdS geometry is
deformed to that of an asymptotically AdS black hole\cite{Witten:1998qj}. The metric is given by
\begin{equation}
ds^2 = \frac{r_H^2 u^2}{L^2} \left[-(1-\frac{1}{u^4})dt^2 + \sum_i^3dx_i^2\right] +
\frac{L^2}{u^2}\left[(1-\frac{1}{u^4})^{-1} du^2 + u^2 d\Omega_5^2 \right],
\end{equation}
where the metric for the 5-sphere can be chosen to be
\begin{equation}
d\Omega_5^2= d\theta^2 + \cos^2\theta d\zeta_1^2+ \sin^2\theta (d\zeta_2^2 + \sin^2\zeta_2 d\zeta_3^2 + \cos^2\zeta_2
d\zeta_4^2),
\end{equation}
The temperature $T$ of the system is related to the radius of the black hole horizon $r_H$ by
\begin{equation}
T = \frac{r_H}{\pi L^2}
\end{equation}

\subsection*{Embedding}

In the D3-D7 setup, the probe wraps $\{x_0, x_1, x_2,x_3,u,\zeta_2,\zeta_3,\zeta_4\}$ while in the D3-D3 setup
the probe wraps around $\{x_0, x_1, u, \zeta_1\}$. We will turn on a world-volume gauge field $A_0(u)$ which
corresponds on the gauge theory side to having a chemical potential or a finite baryon density. In the presence
of a chemical potential, it can be shown that the so-called black hole embedding is a physical embedding, which
we will focus on.
It is common to make a change of coordinates $\cos\theta=\chi(u)$ in the case of D3-D7, and for D3-D3 it is
convenient to pick instead $\sin\theta=\chi(u)$.
For black-hole embeddings, we require regularity at the horizon. In the literature\cite{Myers:2008cj}, this is implemented by switching to a coordinate $\rho= \sqrt{u^2 + \sqrt{u^4-1}}$ and requiring that $\partial_\rho \chi(\rho=1)=0$\footnote{In the near horizon limit $\rho \approx 1 + \sqrt{u-1}$. Therefore requiring $\partial_\rho \chi(\rho=1)=0$ is equivalent to requiring that the expansion of $\chi$ in $u$ to go like $\chi_0 + \chi_1(u-1)+...$. }. We also set $\chi(u=1)=\chi_0$.

The resultant probe brane action for the D3-Dp setup is given by
\begin{eqnarray}\label{freeenergy}
 \frac{\mathcal{F_{\textrm{bulk}}}}{k_B T} &=& I_{\textrm{bulk}} = \mathcal{T_{\textrm{Dp}}}\int d^{p+1}\sigma \sqrt{\det(G+ 2\pi\alpha' F)}\nonumber \\
&=&\frac{1}{k_B T}\int d^{q}\sigma du L_{\textrm{bulk}}\nonumber \\
&=& \frac{\mathcal{N}}{k_B T}\int d^{q}\sigma du (1-\chi^2)^{\frac{p-1}{4}}
\sqrt{|\tilde{g}_{00}|\tilde{g}_{11}^{\frac{p-1}{2}}\tilde{g}_{uu} -(2\pi\alpha')^2\tilde{g}_{11}F_{0u}^2},
\end{eqnarray}
where $\tilde{g}$ is the induced world-volume metric and explicitly for the D3-D3 case, and allowing for a
non-trivial profile in $\chi(u)$, we have
\begin{equation}
d\tilde{s}^2 = \frac{r_H^2 u^2}{L^2} \left[-(1-\frac{1}{u^4})dt^2 + dx^2\right] +
\frac{L^2}{u^2}\left[(1-\frac{1}{u^4})^{-1}(1+ u^2 \frac{(\partial_u \chi)^2}{(1-\chi^2)}) du^2 + u^2
(1-\chi^2)d\zeta_1^2 \right].
\end{equation}
Here, $F = dA$ are world-volume gauge field strengths.
In the Euclidean signature, the integral over the time direction is between zero and $1/(k_B T)= 1/T$ in our units where $k_B$, the Boltzmann constant, is set to one. This factor
is made explicit after the second equality in equation (\ref{freeenergy}).
The normalization constant $\mathcal{N}$ is given by
\begin{equation}
\mathcal{N}= N_0 T^{q+1},\qquad N_0 = \frac{2V_{p-q-1} N_\textrm{c}
N_\textrm{f}\lambda^{\frac{p-3}{4}}}{(2\pi)^{(p-1)}}.
\end{equation}
where $q$ is the spatial dimension of the intersection domain, and $V_{p-q-1}$ is the volume of the sphere
wrapped by the probe D$p$ brane. We choose the
gauge $A_u=0$. The field strengths can be readily solved in terms of $\chi$.
\begin{equation}
F_{0u} =  d \sqrt{\frac{ \tilde{g}_{uu}|\tilde{g}_{00}|(e^2 -
|\tilde{g}_{00}|\tilde{g}_{11})}{\tilde{g}_{11}( b^2 \tilde{g}_{11} -d^2|\tilde{g}_{00}| -
\big(\tilde{g}_{11}(1-\chi^2)\big)^{q}|\tilde{g}_{00}|)}}\,.
\end{equation}
where $b,d$ and $e$ are dimensionful integration constants from solving the equations of motion of
$F_{ux}$,$F_{0u}$ and $F_{0x}$ respectively.
They in fact parametrise the conserved magnetic and electric
charges of the solution. Note that we have absorbed a $1/2\pi\alpha'$ in the definition of the constants $b,d$ and $e$.
For notational simplicity we will take $L=1$, unless stated otherwise.

We will concentrate on the simple case where there is a finite baryon density,
corresponding to a non-trivial $d$, while leaving $b=e=0$.

\begin{equation}\label{asympa}
A_0(u) \sim \bigg\{
\begin{array}{lr}
\mu + \frac{(d/r_H^2 )}{u^{2}}  \hspace{0.1cm} & q=3,\\
\mu + d \log u & q=1.
\end{array}
\end{equation}
near the boundary.
The thermodynamics interpretation of various quantities will be discussed in the next
section. So for the numerics we have two tunable parameters $\chi_0$ and $d$. The near horizon expansion for
$\chi$ is given by

\be \chi=\chi_0+ \chi_1 (u-1)+ O((u-1)^2), \,, \ee with $\displaystyle
\chi_1=-\frac{3\chi_0 (1-\chi_0^2)^3}{4 (\tilde{d}^2+(1-\chi_0^2)^3} $ for D3-D7 and $\displaystyle \chi_1=
-\frac{\chi_0 (1-\chi_0^2)}{4 (\tilde{d}^2+(1-\chi_0^2))}$ for the D3-D3 setup.
Here we have introduced dimensionless $\tilde{d}$ defined as
\begin{equation}
\tilde{d} = \frac{d}{r_H^3} \qquad q=3, \qquad \tilde{d} = \frac{d}{r_H}\qquad q=1.
\end{equation}
Typical profiles for D3-D3 at various
values of $d$ are plotted in figure (\ref{profiles}). \FIGURE[ht]{
\includegraphics[width=0.6 \textwidth]{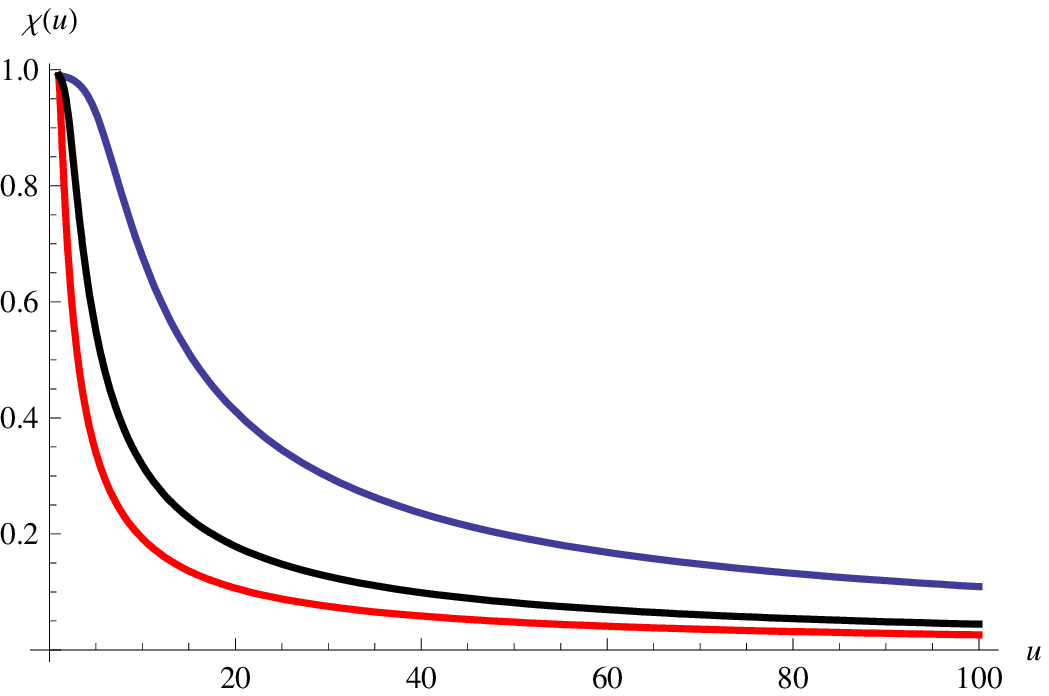}
\caption{The figure plots examples of the embedding $\chi$ at $\tilde{d}=0.065$ (red curve), $\tilde{d}=0.5$ (black curve) and
$\tilde{d}=1.6$ (blue curve).} \label{profiles}}
The asymptotic expansion for $\chi$ reads
\begin{equation}\label{asympchi}
\chi(u) \sim \bigg\{
\begin{array}{lr}
\displaystyle \frac{m}{u} + \frac{c}{u^3}  &  q=3, \\
\displaystyle \frac{m}{u}+\frac{c \log u}{u} & q=1.
\end{array}
\end{equation}

In the supersymmetric limit $m/u$ is an exact solution, both for $q=1$ and $q=3$. Hence, we
will continue to use the same notation $m$ and $c$ for the D3 probe, as in the higher dimensional case.
At $q=1$, $\chi$ saturates the Brietenlohner-Freedman bound. The asymptotic behaviour presented in (\ref{asympchi}) is as expected since
the conformal dimension of the dual operator $\Delta_\pm = (q+1)/2 \pm \sqrt{(q+1)^2/4+ \tilde{m}^2}$, where $\tilde{m}$ is the mass of the scalar field ($\tilde{m}^2=-1$ for $\chi$), evaluates to $\Delta_+ = \Delta_-= 1$ for $\chi$ at $q=1$. This already suggests that the two possible large $r$ behaviour are $\log (r/r_0)/r$ and $1/r$, as we have confirmed. This however, also implies that only one conformally-invariant boundary condition is possible\cite{Klebanov:1999tb}. One could require physical fluctuations to behave as $1/r$ and setting the $\log (r/r_0)/r$ terms to zero.  These, as mentioned above, correspond to the supersymmetric solutions. On the contrary, requiring the vanishing of the $1/r$ term at the boundary and keeping the $\log (r/r_0)/r$ term violates conformal invariance, since that is a scale dependent statement i.e. requires the choice of an arbitrary scale $r_0$. There is therefore likewise only one natural way to incorporate an external source $c_0$, and that is to require that $\chi$ behaves like $c_0 \log(r/r_0)/r$ in the AdS boundary $r\to \infty$.
From the point of view of the dual field theory, this means that the expectation value of the dual operator has a logarithmic violation of conformal invariance in the presence of a source, since the two point function contains a logarithmic divergence. Considerations here apply similarly to the leading $\log$ term of the gauge field in (\ref{asympa}). We will encounter this issue again in the discussion of the thermodynamics of the system.

Therefore, the roles played by $m$ and $c$ are interchanged, namely that $m$ corresponds to the
quark condensate and $c$ to a source\cite{Karch:2005ms}.
This is in contrast with the $q>1$ scenarios \cite{Mateos:2007vn}. We will make further comments about the appearance of the $\log$ as we proceed.




\section{Thermodynamics}
In this section we consider the thermodynamics of the 1+1 defect CFT. There are some interesting differences as
compared to higher dimensional defects as will become clear. These differences arise due to the $\log u$ terms
in the asymptotics of $\chi$ and $A_t$.
We introduce the dimensionful source $\bar{C} = c \pi T$, which is the 1D analog of the quark mass in higher dimensional probes \cite{Mateos:2007vn}, and
\be n_q = \int d\zeta_1 \frac{\partial L_{\textrm{bulk}}}{\partial F_{rt}}= 2 \pi (2\pi\alpha'N_f T_{D3}) d.
\ee
Since $\bar{C}$ and $n_q$ are parameters of the dual defect CFT at $q=1$, the phase diagram of the system is obtained
by keeping them fixed as the temperature $T$ is varied. The ratio $n_q/\overline{C}$ is therefore a handy parameter to label
the dual CFT. We will find that below a certain value for $n_q/\overline{C}$ the system exhibits an instability, similar to the
D3-D7 case. We will be interested
in values of $n_q/\overline{C}$ where the system is stable. It should also be noted that while the Mermin-Wagner theorem states that
at finite temperatures continuous symmetry cannot be generally broken in dimensions $d\le 2$, the restriction is evaded in the large $N$ limit \cite{Hartnoll:2009sz}.


To analyse the thermodynamics of this system we begin by computing the Euclidean action on-shell. From the
asymptotic behaviour of the fields, it is clear that the action is divergent. The counterterms needed to cancel
the divergence arising from $\chi$ have been given in\cite{Karch:2005ms}.
\begin{eqnarray} \label{count1}
L_1 &=& -\frac{\mathcal{N}}{2}\sqrt{\gamma}, \\
L_2&=& \frac{\mathcal{N}}{4}\log (\Lambda/r_0) \sqrt{\gamma}R_{\gamma}, \\
L_4&=& \frac{\mathcal{N}}{2} \sqrt{\gamma} \chi^2(x,\tilde{\Lambda})(1-\frac{1}{\log (\Lambda/r_0)}),
\end{eqnarray}
where $\gamma$ is the induced metric on the boundary, $R_\gamma$ is the Ricci scalar evaluated on $\gamma$, and we have defined $\tilde{\Lambda}=\Lambda/r_H$.
Since we are considering a flat boundary theory, $L_2$ does not contribute. It is important to note here that, since we have switched coordinates to dimensionless $u$, all functions
of $u$ in the boundary limit is evaluated at $\tilde{\Lambda}$. On the other hand, when the cut-off appears explicitly as a coefficient in the counter terms, it appears simply as $\Lambda$, and in the case of explicit $\log$ terms,  $\log(\Lambda/r_0)$, for some arbitrary scale $r_0$.
To make sense of the dependence on $r_0$, one should recall that there is a logarithmic violation of conformal invariance at non- vanishing $d$ and $c$, as discussed in the previous section. Therefore there is dependence on an arbitrary scale $r_0$, which appears inside these $\log$'s. The dependence on $\log r_0$ reminds us that whenever we perform a rescaling in $r$ i.e. $r\to l r$, corresponding to a rescaling to a different energy scale in the dual theory, a contact term (which contributes only to the finite part of the Green's function) proportional to $\log l$ would appear in the action, as discussed in \cite{Bianchi:2001de}. This will not affect the physics (e.g. spectral functions, quasi-frequencies etc) we are interested in and we will set $r_0$ to unity in the rest of our discussions.


For non-vanishing $d$, there is also a logarithmic divergence in the action of the form $-d^2/2 \log
\tilde{\Lambda}$. This is a new divergence only appearing in 1+1 dimensions. We will add

\be \label{count2} L_F= \frac{\mathcal{N}}{2\log\Lambda}A_\mu A_\nu \gamma^{\mu\nu} \sqrt{\gamma}\,, \ee as a
counterterm to remove the logarithmic divergence due to the gauge field. Note that the term preserves gauge
invariance despite looking otherwise. The reason is that AdS/CFT puts restriction on the allowed gauge transformation $\delta A = d\Phi$ such that the asymptotics of  $A_{\mu}$ are not altered. Since $A_{\mu}|_{u\to\infty} \sim a_{0}+ a_{l}\log u$, the leading term at the boundary of $d\Phi$ could at best be $u^{-1}$.
The gauge transformation of (\ref{count2}) again goes like $u^{-1}$ and thus vanishes at the AdS boundary. It is
shown in the appendix that in the gauge $A_u=0$, the boundary term can again be written in terms of gauge invariant variables.

What is the thermodynamics interpretation of $\mu, d, m$ and $c$? Firstly we observe that with a radial electric
field, there is effectively a number $n_q$ of fundamental strings stretching along the worldvolume of the probe
D3 brane. This density of strings is given by $d=-\delta I_{\textrm{bulk}}/\delta F_{t u}$. As a result it is natural to
interpret $d$ as being proportional to a number density. This leads to identifying $\mu$ as the chemical
potential\footnote{As we will see below the physical chemical potential $\overline{\mu}$ differ from $\mu$ by a constant in the case of a probe D3 brane. This applies also to the discussion of $m$, which will be discussed below.}. Similarly in cases of the higher dimensional defects, the brane separation which is controlled by $m$
is identified as being proportional to the
inverse temperature. However, unlike the higher dimensional defects, for the probe D3,
$m$ could be interpreted as the brane separation only in the supersymmetric theory.
Moreover from the point of view of AdS/CFT, the
source of the dual operator should be identified as the coefficient of the $\log u/u$ term i.e. $c$. As
a result we will expect \be \frac{\delta \mathcal{F}}{\delta n_q}=\mu\,, \qquad \frac{\delta \mathcal{F}}{\delta c}=m\,. \ee This is
consistent with the interpretation in \cite{Karch:2005ms} where $m$ is identified as a vev in the
dual field theory. Thermodynamically, we then have that the Euclidean action is a function of $n_q$ and the
source $c$ rather than $\mu$ and $m$ as in the higher dimensional defect case.  Since $\overline{C} = c (\pi T)$ is kept fixed and corresponds to a source in the dual defect CFT, we can identify $1/c$ as the temperature. The free energy $\mathcal{F}$ can thus be interpreted
as the Helmholtz free energy.

The counterterms arising from $L_1+L_2+L_3+L_4$ computes to
\begin{eqnarray}\label{fullcount}
I_{ct}=&&\frac{\mathcal{N}}{k_B T}\int d^q\sigma \bigg(-\frac{\tilde{\Lambda}^2}{2}+ \frac{d^2 \log \tilde{\Lambda}+2 \mu d
}{2r_H^2}+\frac{1}{2}(m + c\log \tilde{\Lambda})^2 - m c \nonumber \\
&&-\frac{c^2}{2}\log \tilde{\Lambda} -\frac{(d^2-r_H^2 c^2)\log r_H}{2r_H^2}\bigg)\,,
\end{eqnarray}
and we define $\mathcal{F}/{k_B T}= I = I_{\textrm{bulk}} + I_{ct}$.
Note that the last term above arises precisely because of the explicit appearance of $\log \Lambda/r_0$ in the definition of the counter terms.

After considering the variation of the action, this leads to \be \delta \mathcal{F}/\mathcal{N}=  -(m- c\log r_H) \delta c+ (\mu- d\log r_H)/r_H \delta \tilde{d}. \ee
This confirms our expectation that the action is a function of $c,\tilde{d}$. It is not surprising that $m- c\log r_H$ and $\mu- d\log r_H$ appears in this particular combination. This is because the dimensionful vev's are defined as
\be
\chi(r)= \frac{1}{r}(\overline{M}+ \overline{C} \log r), \qquad A_r(r) = \overline{\mu} + d \log r,
\ee
implying that $\overline{M} + \overline{C}\log r_H =r_H m$ and similarly $\overline{\mu}+ d\log r_H= \mu$. It is thus consistent to have these
\emph{physical} quantities $\overline{M}$ and $\overline{\mu}$ appearing in the variation of the free energy.



\subsection{Specific heat}

Having discussed the interpretation of the Euclidean action in the previous section, we are ready to compute the
specific heat capacity.

\subsubsection{Case I: $\chi=0$}

For simplicity, let us begin with the massless case where $\chi =0$. The entropy is obtained by differentiating
the action at constant chemical potential. The chemical potential is given by the constant in the asymptotic
expansion of the gauge potential $A_0$.
\begin{equation}
\overline{\mu}= \mu- d\log r_H = \lim_{u\to \infty} A_0(u)- A_0'(u)(u \log u)- d\log r_H,\qquad A_0(u) = \int_{1}^u du F_{u0}.
\end{equation}

At $\chi=0$, they are explicitly,
\begin{equation}
F_{u0}= A_0'(u) = \frac{r_H d}{\sqrt{r_H^2 u^2 + d^2}},\qquad \overline{\mu} =d (\log 2 - \log(r_H + \sqrt{d^2 + r_H^2})).
\end{equation}

Combining with the counter terms, we have
\begin{equation}
\frac{\mathcal{F}}{N_0}= \frac{1}{4} (-2 r_H \sqrt{d^2 + r_H^2} + d^2 (1 + \log 4) - 2 d^2 \log(r_H + \sqrt{d^2 +
r_H^2})).
\end{equation}
As discussed in the previous section, at zero $\chi$, $\mathcal{F}= k_B T I$ is simply a function of $r_H$ and the baryon density
$d$. Therefore the entropy $S$ is given by
\begin{equation}\label{sant}
\frac{S}{N_0} = -\frac{1}{N_0}\frac{\partial \mathcal{F}}{\partial T}\bigg\vert_d = \pi\sqrt{d^2 + (\pi T)^2},
\end{equation}
and the heat capacity $c_v$ is
\begin{equation}\label{cvant}
\frac{c_v}{N_0} = T \frac{\partial (S/N_0)}{\partial T}\bigg\vert_d = \frac{(\pi T)^2}{\sqrt{(\pi T)^2+d^2}}.
\end{equation}
The heat capacity is linear in the temperature at sufficiently high temperatures as expected of a $1+1$
dimensional quantum system. However, the temperature dependence becomes quadratic at low temperatures in the presence of a non-vanishing $d$. Note also that the result is unchanged if we subtract the zero temperature contribution
to the free energy and the chemical potential. Note that in this limit i.e. $\chi=0$, the entropy and subsequently the heat capacity are independent of the arbitrary scale $r_0$, if we are keeping the baryon density fixed.

\subsubsection{Case II: $\chi\ne 0$}
When $\chi$ is non-zero, the above calculations can be done numerically. However, we would then have to take into account the
contribution of $\chi$ and it's corresponding counter terms. We follow the procedure in \cite{Mateos:2007vn} to obtain
an explicit expression for the entropy $S = -\pi \partial_{r_H}\mathcal{F}_H$, keeping the baryon density $d$ and $\bar{C}=r_H c$ fixed.i.e.
\begin{equation}
\partial_{r_H}c = -\frac{c}{r_H}, \qquad \partial_{r_H}\tilde{d} = -\frac{\tilde{d}}{r_H}.
\end{equation}
The entropy thus has five contributions. \be
S= - \pi (S_i + S_{ii} + S_{iii} + S_{iv}+ S_{v}),
\ee
where
\begin{eqnarray}
 S_i &=& (\partial_{r_H}\mathcal{N})  \frac{\mathcal{F}}{\mathcal{N}},\qquad S_{ii}= (\frac{-\tilde{\Lambda}}{r_H}) L_{bulk}(u)\big\vert_{\tilde{\Lambda}},\nonumber \\
 S_{iii}&=& (-\frac{\tilde{\Lambda}}{r_H})\partial_{\tilde{\Lambda}}I_{ct}, \qquad  S_{iv} = (\frac{-1}{r_H})(c \partial_c + \tilde{d}\partial_{\tilde{d}}) \mathcal{F} = \frac{-\mathcal{N}}{r_H}( \tilde{d} \mu/r_H -c m + (\tilde{d}^2-c^2)\log r_H),\nonumber \\
 S_{v} &=& -\mathcal{N}\frac{(\tilde{d}^2-c^2)}{2}\partial_{r_H}\log r_H = -\mathcal{N}\frac{(\tilde{d}^2-c^2)}{2 r_H}.
\end{eqnarray}
We have made used of the fact that $\delta \mathcal{F}/\mathcal{N} = - (m-c\log r_H )\delta c + (\mu/r_H-\tilde{d}\log r_H) \delta \tilde{d}$ in evaluating $S_{iv}$, and $S_{v}$ is the explicit derivative with respect to $r_H$ in the last $(\tilde{d}^2-c^2)/2\log r_H$ term in the counter terms in equation (\ref{fullcount}).

After a tedious but straight forward computation, we finally have
\begin{equation}
S=  - 2\frac{\mathcal{F}}{T} (1 -  \frac{\mathcal{N}\bigg[ (\mu/r_H-\tilde{d}\log (\pi T)) \tilde{d}-c (m-c\log (\pi T))\bigg]}{2\mathcal{F}})+ \mathcal{N}\frac{(\tilde{d}^2-c^2)}{2 T}.
\end{equation}
The corresponding heat capacity is given by
\begin{equation}
c_v = T \partial_T S\bigg\vert_{M,d} = S + \frac{\mathcal{N}}{T^2}\partial_T T^2 ((\mu/r_H-\tilde{d}\log \pi T) \tilde d- (m-c\log \pi T) c)- \mathcal{N}\frac{(\tilde{d}^2-c^2)}{T}.
\end{equation}
Derivative of $c$ and $\mu$ with respect to $T$ has to be implemented numerically.
Typical plots of the free energy against temperature at fixed $d/\overline{C}$ in the stable and unstable regimes are shown in figure (\ref{Cv_T})\footnote{The dotted line is added in by hand. Our numerical results end just before the dotted line begins and the region corresponds to very small $\tilde d\sim 0.01$ and an initial condition for $\chi$ at the horizon extremely close to one $\chi_0 \sim 1-10^{-7}$. It would require much more numerical accuracy to see the dotted line.  }.
\FIGURE[ht]{\label{Cv_T}
\begin{tabular}[ht]{cc}
\includegraphics[width=0.45\textwidth]{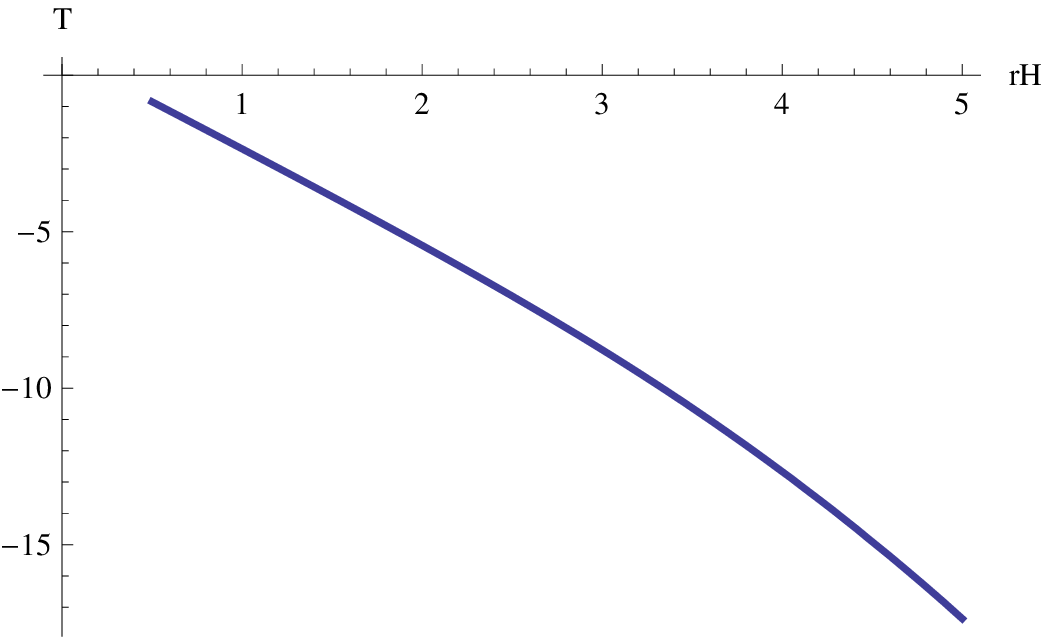} &\includegraphics[width=0.45\textwidth]{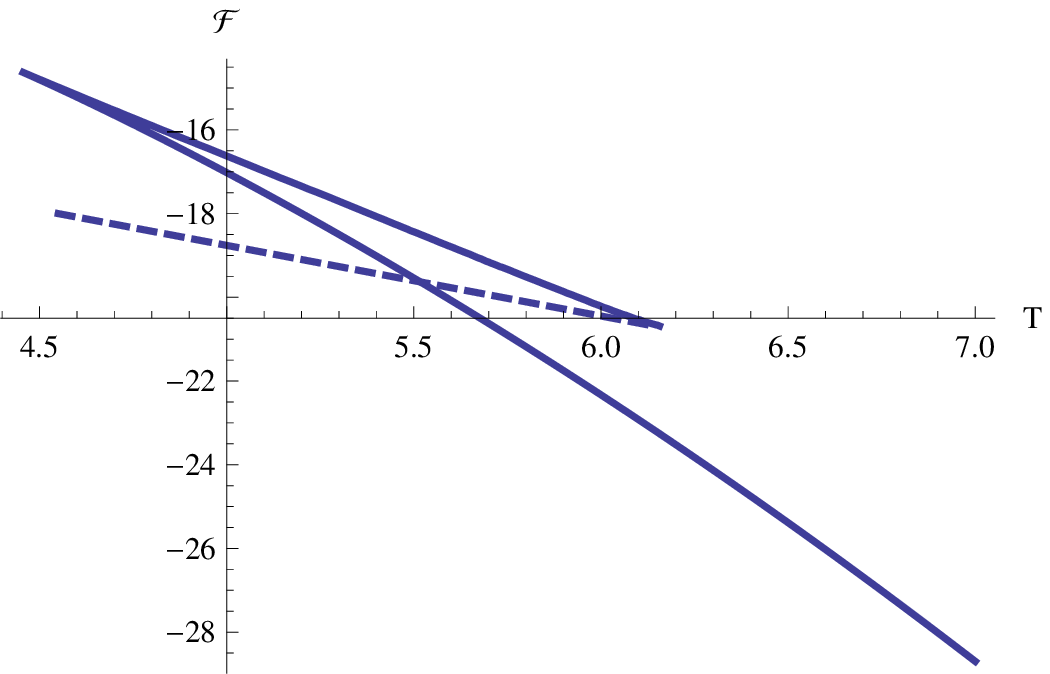} \\
{(a)} &  {(b)}\\
\includegraphics[width=0.45\textwidth]{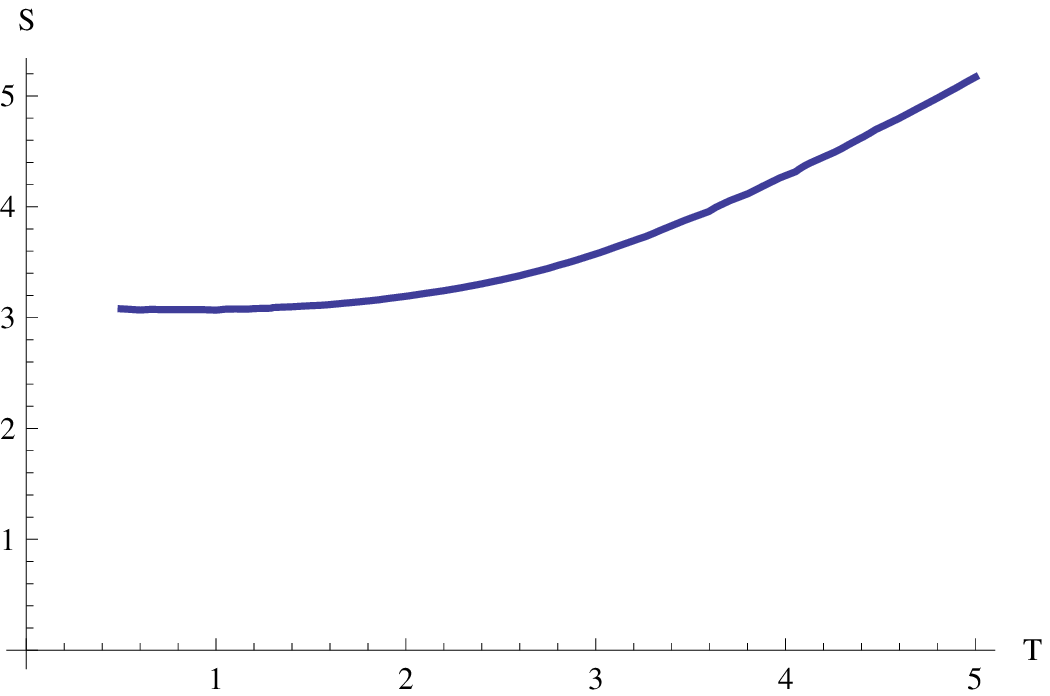} &\includegraphics[width=0.45\textwidth]{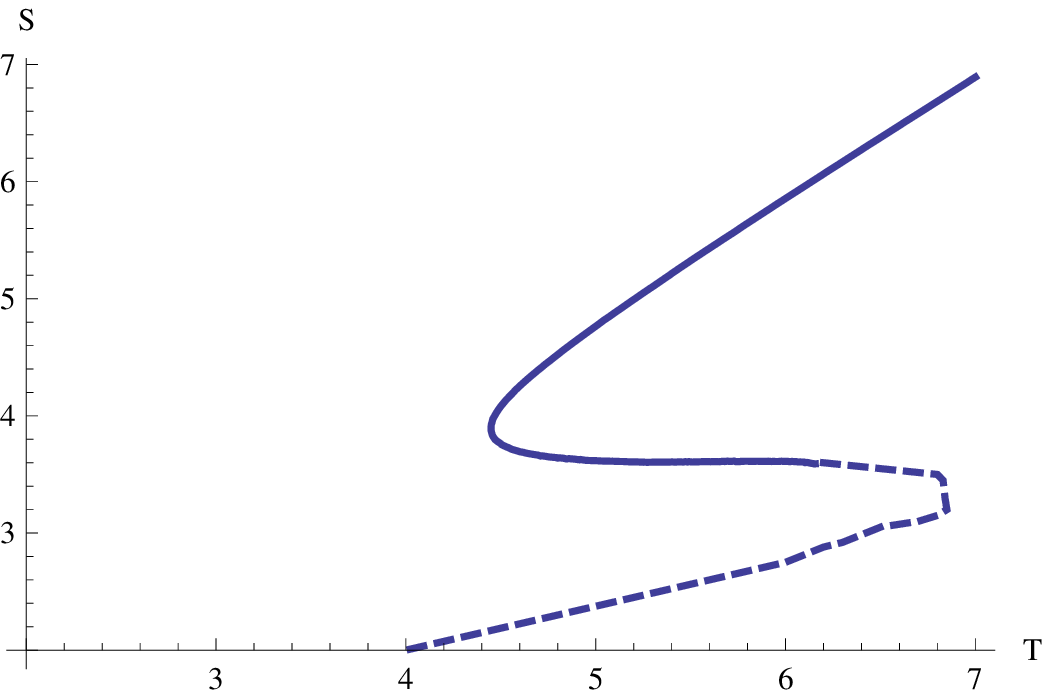} \\
{(c)} &  {(d)}\\
\end{tabular}
\caption{(a) and (c) are plots of free energy and the corresponding entropy against temperature in the stable regime, where $d/\bar{C}>1$. (b) and (d) are typical plots of $\mathcal{F}$ and $S$ in the unstable regime, where $d/\bar{C}<1$. The dotted line is added in by hand.}}
At high temperatures, both the entropy and the specific heat scales as $T$, approaching the conformal result.
Below $d/\bar{C}<1$, the heat capacity could turn negative. This allows us to determine the phase diagram of the system.
This agrees with a quasi-normal mode analysis, where an unstable mode that grows in time appears
where the heat capacity is negative. The phase diagram is shown in figure (\ref{phase_plot})\footnote{At $T=0,d=0$, the supersymmetric solution $\chi=m/u$ is stable. Note that one could not smoothly go from a general black hole embedding to a supersymmetric solution by taking the $c\to 0$ limit. For more general $\chi$,the Minkowski embedding is the only stable embedding \cite{Mateos:2007vn}.}.

\FIGURE[ht]{\label{phase_plot}
\includegraphics[width=0.6\textwidth]{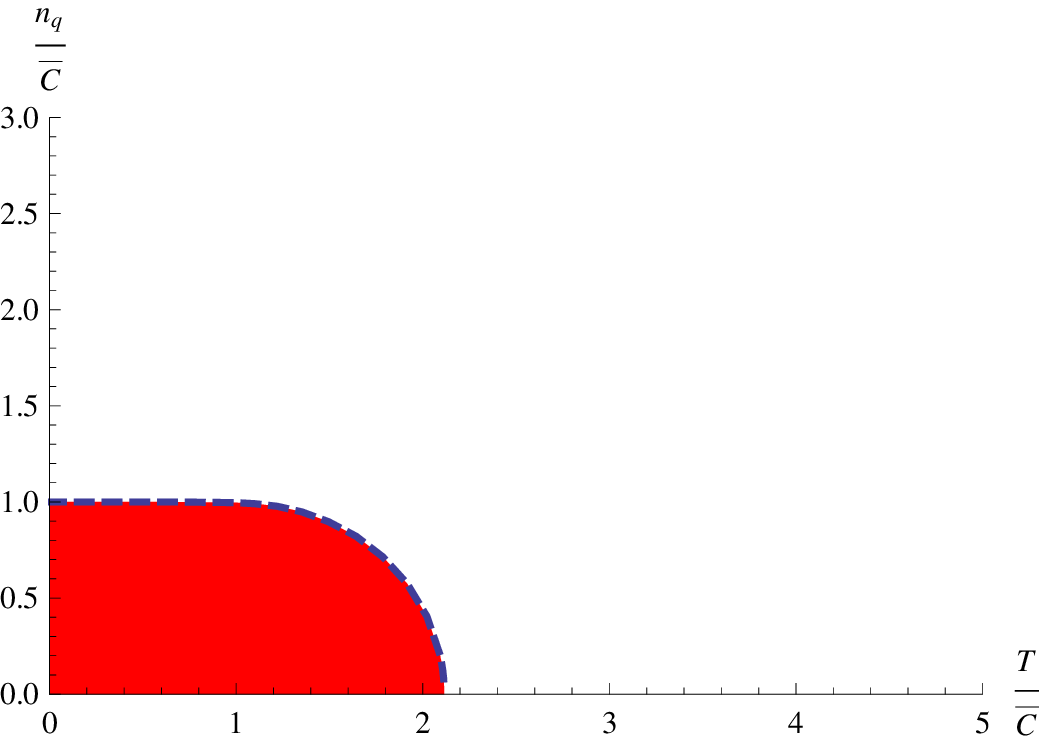}
\caption{The phase diagram of the system. The shaded region (excluding the origin) is unstable.}}




\section{Sound modes}\label{soundmodes}

For simplicity, we would like to begin by considering the simpler embedding where $\chi=0$ all the way. The terms in the
action mixing the gauge potential and the brane profile perturbations are zero in this case. Therefore we can
consider the equations of motion of the longitudinal gauge perturbations independently of the other scalar
fluctuations. The calculation in the zero temperature limit mimics that in \cite{Karch:2008fa}. The solution
there however cannot be applied directly for probe D3's by simply substituting the defect dimension $q=1$ in our
case. The probe D3 case however is straightforward to solve and one obtains a dissipationless mode with a
dispersion given by the conformal value $\omega=\pm k$. In the finite temperature, the near horizon solution has
a singularity and requires slightly more work. It is convenient to work with the coordinate $z=1/u$, which has the correct normalisation at the boundary and would avoid mistake when we begin to expand in power series of $\omega$ and $q$.
The equation of motion at zero $\chi$ is given by

\begin{eqnarray}\label{sound}
&& \partial_z (F[z] E_x'[z]) + G[z] E_x[z] =0, \nonumber \\
&& F[z] = \frac{z (1 + \tilde{d}^2 z^2)^{3/2} (-1 + z^4)\omega}{2 q^2 (-1 + z^4) + 2 (1 + \tilde{d}^2 z^2) \omega^2}, \nonumber \\
&& G[z] =  \frac{z^3 \sqrt{1 + \tilde{d}^2 z^2}}{2 (-1 + z^4)}.
\end{eqnarray}


We would like to solve
these equations in the limit, where $\omega$ and momentum $k$ are much less than any other scales,
in which case we can obtain a solution perturbatively in $\omega$ and $k$. We will follow the approach in
\cite{Nunez:2003eq}. To begin with, we isolate the singularity appearing in the near horizon limit, and write
\begin{equation}
E_x[z] = (1-z)^{\frac{-i \omega}{4}}(e_0[z] + \omega e_1[z]+ ...)
\end{equation}
After making the substitution in equation (\ref{sound}) and removing the overall factor of $(1-z)^{\frac{-i
\omega}{4}}$, we expand the resulting equations of motion in powers of $\omega$ and $k$. In the expansion we
considered the case $\omega \sim k$.
For each solution of $e_0,e_1...$ we have to set two boundary conditions. Since the singular part of the
solution has been isolated, we require that $e_0[z]$ and $e_1[z]$ and so on, are regular at the horizon. Also,
by convention we choose to absorb any constants $e_{i>0}[z \to 1]$ by $e_0$, and therefore require that
$e_{i>0}[z]$ vanish at the horizon. Given these conditions, we find that
\begin{eqnarray}
e_0[z] &=& c_0 \nonumber \\
e_1[z] &=& c_0(c_1 - \frac{i (k^2 - \omega^2) \log[z]}{\omega^2}+ \textrm{other terms})\end{eqnarray} The solution
$e_1[z]$ involves other complicated terms in $\log[1+\tilde{d}^2z^2]$ and $\log[1+\tilde{d}^2z^2 + \sqrt{(1+\tilde{d}^2)(1+\tilde{d}^2z^2)}]$,
which are regular at the boundary and do not concern us here. The constant $c_1$
can be expressed in terms of $\tilde{d}, \omega$ and $k$, by requiring that $e_1$ vanishes at the horizon. What is
important however, is that the coefficient of the $\log[z]$ term is given by $(k^2- \omega^2)$, which gives a
quasi-normal mode at
\begin{equation}
\omega = \pm k,
\end{equation}
in agreement with the conformal result of a sound mode in one-dimension. We can in fact evaluate the spectral
function explicitly. To lowest order in $\omega$ and $k$, it is given by
\begin{equation}
\frac{\textrm{Im}(G_{xx})}{(2 \mathcal{N})/(\pi^2 T^2)}= \frac{\omega^2}{\omega^2-k^2} \textrm{Im}(\frac{c_0}{-\frac{i c_0(k^2 -
\omega^2)}{\omega}}) =\frac{\omega^3}{(\omega^2- k^2)^2},
\end{equation}
and similarly for other components of the Green's functions of the currents, which are obtained by putting into
the above expression appropriate pre-factors of $k$'s and $\omega$'s.

The real part of $G_{xx}$ approaches a finite value in the hydrodynamic limit at $k=0$. This is because as we
consider higher order terms we find
\begin{eqnarray}
\lim_{\omega\to 0}\textrm{Re}(G_{xx}) &\sim& \textrm{Re}(\frac{1+  \omega (H_R+ i H_I ) + ...}{i \omega (1+ \omega (K_R + i K_I)+...)}) \nonumber \\
&=& H_I - K_I
\end{eqnarray}
where $H_{R,I}$ and $K_{R,I}$ are overall zeroth order constants depending on $\omega,k$. We will have to solve
for still higher order corrections in order to determine $K_I$.

We followed this sound mode using the complete equation {\it without} taking the hydrodynamic limit and found
that this dispersion relation remains exact beyond the hydrodynamic limit. In fact, even when $\chi$ is non-zero
and mixing becomes important, as we will investigate in the following sections, this mode remains intact and
dissipationless.

This should be put in sharp contrast with higher dimensional defect theories obtained from higher dimensional
probes (i.e. D5 or D7), where the leading term in the boundary expansion of the gauge field is given by the
constant term. However, for regularity at the horizon, $e_0$ continues to be a constant in these higher
dimensional theories. The constant term at the boundary is then given by
\begin{equation}
E_x[z] = c_0(1 + \omega (c_1) + ...)
\end{equation}
where $c_0$ is an arbitrary constant, and $c_1$ depends on $k$ and $\omega$ although it is overall zeroth order
in $\omega$ and $k$. The quasi-normal modes are defined at the zeroes of the constant term. It is however clear
that $\omega$ cannot begin at linear order in $k$, due to the presence of the constant term $c_0$. This absence
of a sound mode has been observed in \cite{Kim:2008bv}, and is interpreted as the destruction of the
Fermi-surface at finite temperature in the strongly coupled regime.

We have also looked for quasi-frequencies of the $\Theta$ fluctuations in the hydrodynamic limit, and for
simplicity, at zero $\chi$ and zero chemical potential. In that case the regular solution zeroth order in
$\omega$ is non-trivial and contains a $\log$ term at the boundary, whose coefficient is independent of $\omega$.
The quasi-normal mode occurs at the zeroes of the coefficient of the $\log$ term, and therefore no higher order
corrections in $\omega$ and $k$ could cancel the zeroth order contribution, if we assume that $\omega \sim k$.
We conclude that it does not contain a {\it sound mode} with linear dispersion.

While the above expression for the quasinormal frequency appears to be independent of $\tilde{d}$ that controls the
chemical potential, the sound mode actually disappears at zero temperature if we then also put $\tilde{d}=0$. The limit
to $\tilde{d}=0$ is thus not a smooth one. The difference lies in the modified near horizon (i.e. $u\to 0$ or $z\to
\infty$) expansion of the solution. Instead of $\exp(\pm i \omega z)/z$ we have $\exp(\pm i z
\sqrt{\omega^2-k^2})/\sqrt{z}$. We pick the plus sign for regularity as in\cite{Karch:2008fa}. The
equations of motion at $\tilde{d}=0$ and $T=0$ can in fact be solved exactly. The solution is given by
\begin{equation}
E_x(\omega, z)= c_1 I_{0}(z\sqrt{k^2-\omega^2})+ \frac{c_2}{\sqrt{\pi}}
K_{0}(z \sqrt{k^2-\omega^2}),
\end{equation}
where $I$ and $K$ are the modified Bessel functions. The constants $c_1$ and $c_2$ are fixed by the boundary condition, and they are related by
\begin{equation}
c_2 = \frac{-i c_1}{\sqrt{\pi}}.
\end{equation}
Then we can expand the solution in the boundary limit, and extract the coefficient of the log term and the
constant. They are given by
\begin{equation}
E_x(z \to 0) = i \gamma + \pi + i \log(\frac{i \sqrt{\omega^2-k^2}}{2}) + i \log(z) + ...
\end{equation}
where $\gamma$ is the Euler's constant. The log term becomes independent of $\omega$ and $k$ and the massless
sound mode disappears.

We can also obtain the Green's function from the solution\footnote{We obtained the results at $ r_H=1$. As we will show more
explicitly in the appendix, the explicit $\log r_H$ contribution in the counter term contributes only to the real part of the Green's function
and does not affect the imaginary part, which is of more interest for our purpose.}. At zero $k$ this simply gives
\begin{equation}\label{gxx0td}
G_{xx} = \frac{2 \mathcal{N}}{\pi^2 T^2}\left(-\mathcal{\gamma} + i \pi - \log(\frac{i \omega}{2})\right).
\end{equation}
A plot of the real and imaginary parts of the Green's function are shown in figure (\ref{G0T0D}). The real part depends
logorithmically on $\omega$, where as the imaginary part is simply a constant $\pi/2$.

\FIGURE[ht]{
\includegraphics[width=0.6 \textwidth]{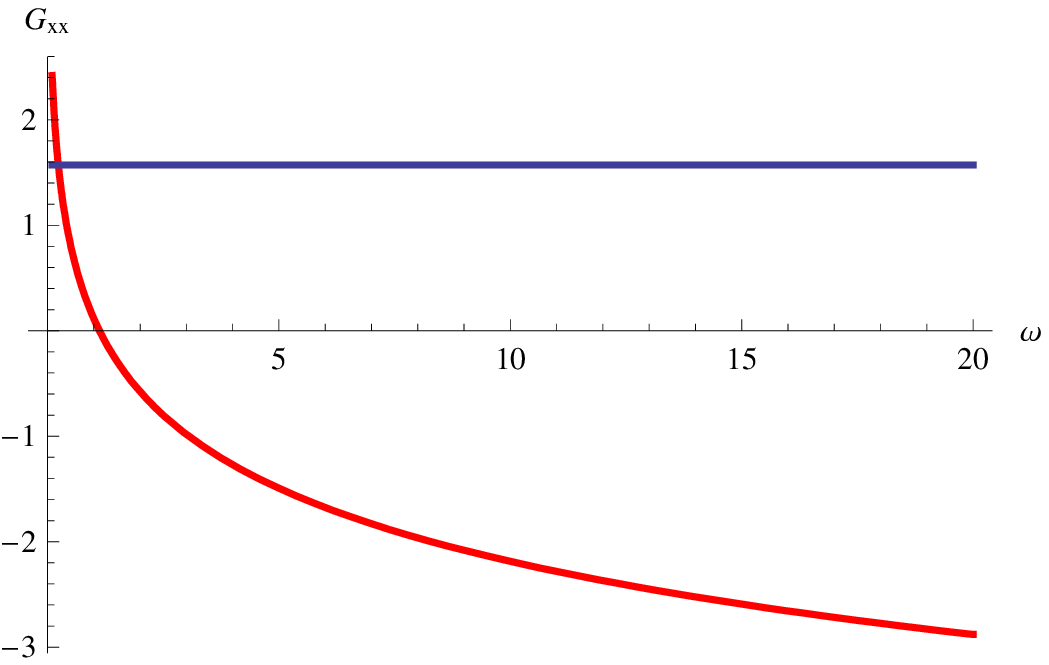}
\caption{The red and blue curves plot respectively the real and imaginary parts of $G_{xx}$ at zero $\tilde{d}$, $T$
and $k$. The imaginary part is a constant given by $\pi/2$. } \label{G0T0D}}

The result also agrees precisely with the high $\omega$ limit of the numerical solution at finite baryon density and temperature, which we will return  to later along with the numerical solutions.

\section{Bosonic fluctuations and spectral functions}
\subsection{The diagonal modes}
Consider the fluctuations\footnote{Note that we have not considered fluctuations of the probe along $x^2, x^3$ inside the $AdS_5$. The action of these modes include contributions from the WZ terms\cite{erd,Gomis:2007fi,Drukker:2008wr}, and the boundary action evaluates to zero. They are related to surface operators in the dual theory\cite{Gomis:2007fi,Drukker:2008wr}. The usual mesonic excitations in the dCFT are described by fluctuations of the gauge field and of the transverse 3-sphere\cite{Mateos:2007vn}. We will not discuss these surface operators any further in this work.} of the angular position $\zeta_2$\footnote{We have not considered fluctuations of other angular directions on the transverse 3-sphere, since they are related to the $\zeta_2$ mode by $SO(4)$ rotations.}, as defined in section (\ref{braneconfig}) of the
probe D3 brane. The diagonal modes exhibit the usual quasi-particle peaks and indicate no sign of the underlying
instability in the system. Here we plot some representative spectral functions for the D3-D3 setup in figure
(\ref{zetaGreen}), since it has never been studied before. The D3-D7 case was studied for example in
\cite{Myers:2008cj}. The first peak is located precisely at $\omega = k$. The width of the peak increases with
$k$, and the Green's function has a singularity at precisely $\omega =k=0$.

\FIGURE[!ht]{
\begin{tabular}[ht]{cc}
\tiny{(a)} & \tiny{(b)}\\
\includegraphics[width=0.45\textwidth]{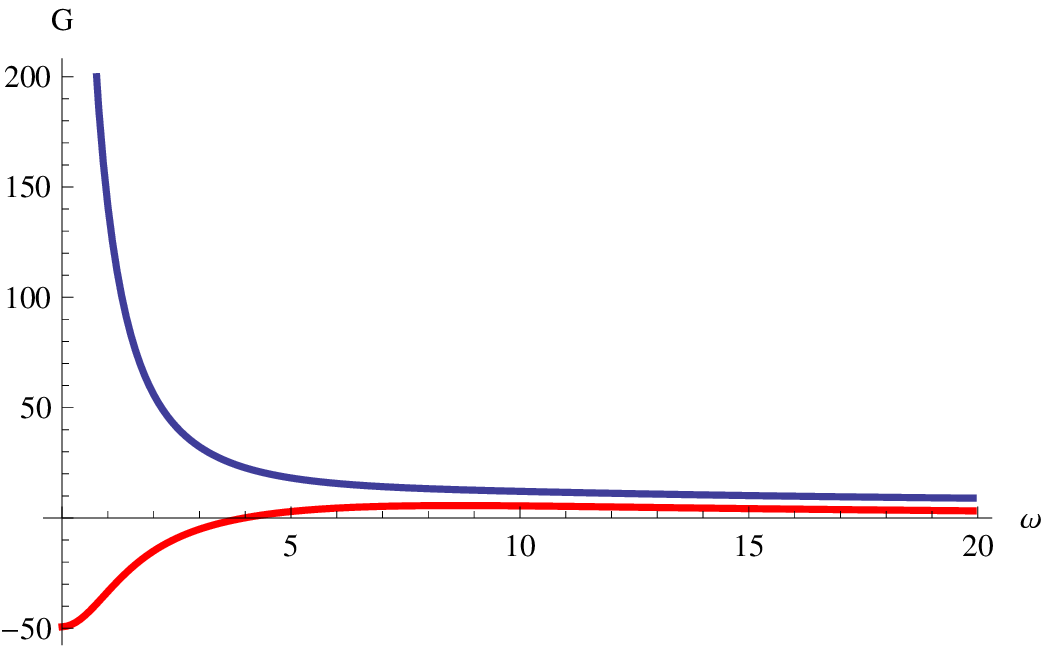}& \includegraphics[width=0.45\textwidth]{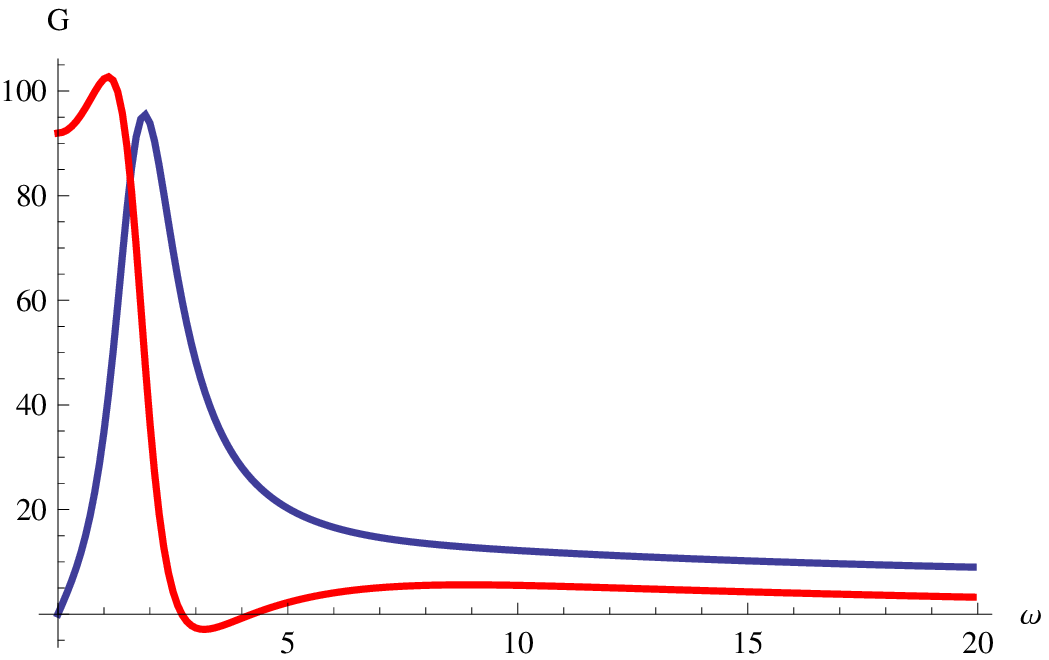}\\
\end{tabular}
\caption{The real (red curve) and imaginary (blue curve)parts of the Green's functions corresponding to
$\zeta_2$ fluctuations, at $r_H=1$, $d/\bar{C}=1.074$ and (a)$k=0$, and (b) at $k=2$. } \label{zetaGreen}}


\subsection{The mixed modes}

In the presence of a chemical potential and at a finite momentum, there is a mixing between the fluctuations of
the longitudinal electric field $E_x$ and the profile $\delta\theta$. This was first pointed out in
\cite{Myers:2008cj}. These mixed modes were subsequently studied in detail in \cite{Mas:2008jz} for the case of probe D7's. This Lagrangian at
quadratic order can be written as

\begin{eqnarray}\label{mixedlang}
{\cal L}&=&T_1 \delta\theta^2+T_2 \delta\theta_x^2 +T_3 \delta\theta_t^2+T_4 \delta\theta_u^2+T_5 \delta \theta\delta\theta_u \nonumber\\
&+& D_1 \delta \theta_x F_{xt} +D_2 \delta\theta_u F_{tu} +D_3 \delta\theta F_{tu} \nonumber \\
&+& S_1 F_{xt}^2 +S_2 F_{tu}^2 +S_3 F_{xu}^2\,.
\end{eqnarray}
To avoid cumbersome notations, we have implicitly made a change of coordinates i.e. $(t,x)\to \frac{1}{\pi T}(t,x)$ . As a result our 2-momenta are dimensionless
quantitites normalized by the temperature.
Here $S_i,T_i, D_i$ are complicated functions of $u$ whose complete expressions we will refrain from showing here. Their near horizon and boundary behaviour will be given in the appendix. The second line denotes mixing terms and all the mixing terms are
proportional to $\tilde{d}$. We will work in the gauge $A_u=0$. This leads to the constraint equation \be\label{cons}
\partial_t D_2 \delta\theta_u + \partial_t D_3 \delta \theta=2 \partial_t (S_2 \partial_u A_t)+2 \partial_x (S_3 \partial_u A_x)\,.
\ee
The equation of motion for $A_x$ and $A_t$ are
\be \label{axeom}
\partial_t (D_1 \delta \theta_x)=-2 \partial_t (S_1 F_{xt})+2 \partial_u (S_3 \partial_u A_x)\,,
\ee
\be \label{ateom} -\partial_x (D_1\delta\theta_x)+\partial_u (D_2 \delta\theta_u)+\partial_u(D_3\delta\theta)=2
\partial_x (S_1 F_{xt})+2 \partial_u (S_2 \partial_u A_t) \,. \ee Now we write
\begin{eqnarray}
A_\mu &=& \int \frac{d\omega dk}{(2\pi)^2} e^{-i \omega t +i k x} a_\mu(u,\kappa)\,,\\
\delta\theta &=& \int \frac{d\omega dk}{(2\pi)^2} e^{-i \omega' t +i \kappa' x} \Theta(u,\kappa)\,.
\end{eqnarray}
Since $A_\mu$ and $\delta\theta$ are real, we have \be \bar\Theta_{-\kappa}\equiv\bar\Theta(u,-\kappa)=\Theta(u,\kappa)\equiv
\Theta_\kappa\,,\qquad \bar a_{\mu,-\kappa}\equiv\bar a_\mu(u,-\kappa)=a_\mu(u,\kappa)\equiv a_{\mu,\kappa}\,, \ee where $~\bar{}~$
denotes complex conjugation. The gauge invariant combination is $E_x=k a_t+\omega a_x$. Using the constraint
equation we find \be \label{axEx} a_x'= {k \Delta +2 i \omega S_2 E_x'\over 2 i (\omega^2 S_2+k^2 S_3)}\,, \ee
or equivalently \be \label{atEx} a_t'={2 i k S_3 E_x'-\omega \Delta\over 2 i (\omega^2 S_2+k^2 S_3)}\,, \ee
where \be \Delta= -i \omega D_2 \Theta'- i\omega D_3 \Theta\,. \ee Using this we find \be D_1 \omega k \Theta=-2
S_1 \omega E_x +2 \partial_u \left(S_3 {k \Delta +2 i \omega S_2 E_x'\over 2 i (\omega^2 S_2+k^2 S_3)}\right)\,.
\ee After some algebra, the equation of motion for $\theta$ is found to be \be \label{thetaeom} 2 \partial_x
(T_2 \delta\theta_x+{1\over 2} D_1 F_{xt})+2 \partial_t (T_3 \delta\theta_t)+2 \partial_u
(T_4\delta\theta_u+{1\over 2} T_5\delta\theta)-2 T_1\delta\theta -T_5\delta\theta_u+D_3 A_t'-\partial_u (D_2
A_t')=0\,. \ee

Note that our equations of motion explicitly violates spatial and time reversal. This
is a result of the presence of a background worldvolume electric field $F_{ut}$. This will have implications in
the spectral functions, which we will discuss in the next sub-section.



\subsubsection{The spectral function of the mixed modes}\label{mixedspectral}

Recall that for the mixed fluctuations, we have two coupled linear second-order differential equations, and as a
result, there should be four independent pair of solutions $(E_x^i(u),\Theta^i(u)), i\in\{1,...4\}$. By imposing
the infalling boundary conditions at the horizon for both $E_x$ and $\Theta$, we are imposing two constraints
and should be left with two independent solutions. The boundary values of $E_x$ and $\Theta$ respectively source
different operators, and given these two independent solutions, it is in principle possible to construct
solutions with independent arbitrary boundary values $E_x(\infty)$ and $\Theta(\infty)$. i.e. in general

\begin{equation}
E_x(u) = A E_1(u) + B E_2(u), \qquad \Theta(u)= A \Theta_1(u) + B \Theta_2(u),
\end{equation}
which allow us to solve for $A$ and $B$ for any given pair of boundary values $\{E_x(\infty),\Theta(\infty)\}$.
The Green's function is obtained by differentiating the on-shell action evaluated at the boundary with respect
to $E_x(\infty)$ and $\Theta(\infty)$. It is therefore apparent that while explicit mixing terms in the on-shell
action vanishes at the boundary, mixing terms involving the product of $\Theta(\infty)E_x(\infty)$ could still
appear through the squares and products of $A$ and $B$, which generally depend on both of these boundary
quantities. It is important to emphasise that the sources $\{E_x(\infty),\Theta(\infty)\}$ are only independent if we have
the freedom to pick $A$ and $B$ accordingly. Given a particular boundary condition at the horizon, the boundary
values of the pair solution $(E_x(z), \Theta(z))$ of the equations of motion are not independent and therefore
evaluating the action at such a particular solution and differentiating with respect to these correlated
boundary values would not give the correct Green's function, as in \cite{Mas:2008jz}. In fact, it is apparent
from the procedure taken in \cite{Mas:2008jz} that the diagonal elements of the spectral function (i.e.
imaginary parts of the Green's function) are not proportional to the conserved current discussed in section
(\ref{conservedj}). By picking different horizon boundary values these spectral functions could turn negative,
which already suggests that the procedure is pathological.


The study of mixed modes is well known in the context of shear modes in R-charged backgrounds\cite{Son:2006em}. The
numerical procedure needed to search for quasi-normal modes have been discussed in \cite{Amado:2009ts}. We will
discuss this procedure, and lay out the explicit way of computing spectral functions\footnote{We thank Andrei Starinets and
Sean Hartnoll for discussion of the issue and pointing us to useful references. }.
To implement the procedure numerically, where we are only capable of controlling the horizon values, we
construct general solutions of $E_x$ and $\Theta$ in the following manner. First, we construct two independent
sets of solutions by picking two sets of horizon boundary conditions, for concreteness, say
\begin{eqnarray}
E_1(u\to 1) \sim (1-u)^{\frac{-i\omega}{4}}(1+ ...) &\qquad& \Theta_1(u\to 1)  \sim (1-u)^{\frac{-i\omega}{4}}(1+ ...),\nonumber \\
E_2(u\to 1) \sim -(1-u)^{\frac{-i\omega}{4}}(1+ ...) &\qquad& \Theta_2(u\to 1) \sim  (1-u)^{\frac{-i\omega}{4}}(1+ ...),\label{bcmix}
\end{eqnarray}
where the ellipses again denote higher order terms in $(u-1)$, which are solved in terms of the lower terms.
Then we define two pairs of solutions
\begin{eqnarray}
e_x&=& E_1(u,\kappa)-{\Theta_1(u_\infty,\kappa)\over \Theta_2(u_\infty,\kappa)}E_2(u,\kappa),\qquad e_\theta= \Theta_1(u,\kappa)-{\Theta_1(u_\infty,\kappa)\over \Theta_2(u_\infty,\kappa)}\Theta_2(u,\kappa)\,,\\
t_x&=& E_1(u,\kappa)-{E_1(u_\infty,\kappa)\over E_2(u_\infty,\kappa)}E_2(u,\kappa),\qquad t_\theta=
\Theta_1(u,\kappa)-{E_1(u_\infty,\kappa)\over E_2(u_\infty,\kappa)}\Theta_2(u,\kappa)\,, \label{constructsol}
\end{eqnarray}
where we denote $\kappa = (\omega, k)$.
These solutions are constructed such that the leading boundary terms in $e_\theta$ and $t_x$ vanish. As a
result, a general solution can be written as
\begin{eqnarray}
E_x&=&n_e \left( E_0(\kappa) {e_x(\kappa,u)\over e_x (\kappa,u_\infty)}+T_0(\kappa) {n_\theta\over n_e}{t_x(\kappa,u)\over t_\theta(\kappa,u_\infty)}\right)\,,\label{mixsole}\\
\Theta &=& n_\theta \left( E_0(\kappa) {n_e\over n_\theta} {e_\theta(\kappa,u)\over e_x (\kappa,u_\infty)}+T_0(\kappa)
{t_\theta(\kappa,u)\over t_\theta(\kappa,u_\infty)}\right)\,,\label{mixsolt}
\end{eqnarray}
 where for D3-D3{\footnote{for D3-D7 $n_e=1\,, n_\theta={1\over u_\infty}\,.$}}
\be n_e=\ln u_\infty\,,\qquad n_\theta={\ln u_\infty \over u_\infty}\,, \ee the normalizations chosen to respect
the asymptotic boundary conditions. $E_0$ and $T_0$ determine the boundary values of $E_x$ and $\Theta$
respectively, and source the corresponding dual operators. The boundary action can be shown to be schematically
\be -\mathcal{N}\int d\omega dk \frac{S_2}{\omega^2-k^2} E_x E_x' +T_4 \Theta \Theta'\,, \ee with which we can compute a matrix of Green functions: \be
G=\mathcal{N}\begin{pmatrix}\displaystyle
2 S_2 n_e^2 {e_x'(\kappa,u)\over e_x (\kappa,u_\infty)} & S_2 n_e n_\theta {t_x'(\kappa,u)\over t_\theta(\kappa,u_\infty)}+T_4 n_\theta n_e {e'_\theta(-\kappa,u)\over e_x(-\kappa,u_\infty)} \\
S_2 n_e n_\theta {t_x'(-\kappa,u)\over t_\theta(-\kappa,u_\infty)}+ T_4 n_\theta n_e {e'_\theta(\kappa,u)\over e_x(\kappa,u_\infty)}
& 2 T_4 n_\theta^2 {t'_\theta(\kappa,u)\over t_\theta(\kappa,u_\infty)},
\end{pmatrix}
\ee
where we have abosrbed the factors of $\omega^2-k^2$ inside $e_x$ to avoid clumsy notations.
We also need to include the contribution due to the counter terms\footnote{Again we are setting $r_H=1$.}. When the divergences are removed the Green's
function would be given by \be\label{Greensmix} G= \mathcal{N}\begin{pmatrix}\displaystyle
{2e^{(1)}_x(k)\over e^{(0)}_x(k)} & {t^{(1)}_x(k)\over t^{(0)}_\theta(k)}+{e^{(1)}_\theta(-k)\over e^{(0)}_x(-k)} \\
{t^{(1)}_x(-k)\over t^{(0)}_\theta(-k)}+{e^{(1)}_\theta(k)\over e^{(0)}_x(k)} & {2 t^{(1)}_\theta(k)\over
t^{(0)}_\theta(k)}
\end{pmatrix}
\ee where we denote with superscript $(0)$ for the coefficients of the leading boundary term and $(1)$ for the
coefficients of the first sub-leading boundary term in the boundary expansion of the solutions. Clearly, the
vanishing of the leading boundary term, corresponding to the defining condition of the quasi-normal modes, again
coincides with singularities of the Green's function matrix.
It is important to note that in (\ref{constructsol}), by construction , the zeroes of $e^{(0)}$ and $t^{(0)}$
coincide, and are proportional to
\begin{equation}
\textrm{QNM}=E_1 \Theta_2 - E_2 \Theta_1.
\end{equation}
This is the same definition for quasi-normal modes as in \cite{Amado:2009ts}. This also implies that all the
entries in the Green's function matrix have a singularity at the same time. This is well known from the study of
the coupled shear modes from the fluctuations of the background metric\cite{Son:2006em}. We will work out the Green's function
matrix explicitly for the case of D3-probes in the appendix.

\subsubsection{Symmetries of the action and the spectral function}\label{conservedj}
In the absence of mixing, we have $$\Theta_\kappa\rightarrow e^{i\phi_{1,\kappa}}\Theta_\kappa\,,\qquad a_{\mu,\kappa}\rightarrow
e^{i\phi_{2,\kappa}} a_{\mu,\kappa}$$ to be symmetries of the action. This leads to a $U(1)\times U(1)$ symmetry for each
$k$ in the bulk action which leads to two (infinite sets of)conserved currents $J_{\Theta_\kappa}$ and $J_{E,\kappa}$.
(This is probably an accidental symmetry of the quadratic action, to be broken by higher terms.)

In the presence of mixing, only $\phi_1=\phi_2$ leads to a symmetry of the action. This happens when one turns
on a non-zero $d$ and a non-zero $k$. In this circumstance, the global symmetry breaks to a $U(1)$ with a single
conserved current. This works out to be
\be\label{current}
J(u)=2 {\rm Im~} \bigg[ T_4 \bar\Theta_\kappa'\Theta_\kappa -
\half D_2\bar\Theta_\kappa'a_{t,\kappa} -\half D_3\bar\Theta_\kappa a_{t,\kappa}+ \half D_2\bar\Theta_\kappa a'_{t,\kappa}+S_3 \bar a_{x,\kappa}'
a_{x,\kappa}+S_2 \bar a_{t,\kappa}' a_{t,\kappa} \bigg]\,.
\ee
Note that terms containing $T_5$ are not present in the conserved current since it is $\sim |\Theta_\kappa|^2$ which is real and is canceled out.
Thus, we have a conservation equation \be
\partial_u J(u)=0\,,
\ee
which leads to a condition for checking our numerics. As noted in \cite{Nunez:2003eq}, the imaginary part of
the boundary terms of the Lagrangian evaluated on shell is proportional to the conserved current (\ref{current}).i.e.
\be\label{bL}
\frac{1}{2} J(u\to \infty)\,\nonumber \\
\ee  and is thus independent of the radial coordinate
$u$.  Also we should note that terms containing $D_i$'s (i.e. all the mixed terms) in fact asymptote to zero at the boundary.

Explicitly, the current evaluated on-shell at the horizon, is given by
\begin{equation}\label{horizoncurrent}
J(1)= \sqrt{1 + \tilde{d}^2 - \chi_0^2} (\omega + \frac{1}{\omega}),
\end{equation}
where we have made use of the near horizon solutions of $\Theta$ and $F_{tx}$ that satisfy the infalling
boundary conditions as given in (\ref{bcmix}). The result is independent of the choice of signs in
(\ref{bcmix}).

The first term in (\ref{horizoncurrent}) originates from the $\bar{\Theta}\Theta'$ term while the second term
from $\bar{a}_{x}a_{x}'$ and $\bar{a}_{t}a_{t}'$. The cross terms again vanish at the horizon. The sign of the
expression is explicitly positive definite, and this follows from the signs of $T_4,S_2$ and $S_3$, and also the
infalling boundary condition. In the absence of coupling, $U(1)\times U(1)$ symmetry is restored, such that the
two terms in $J(1)$ are conserved independently.

Recall that the spectral function are obtained from the quadratic boundary action in eq. (\ref{bL}). In fact,
from our discussion in (\ref{Greensmix}), the imaginary part of each of the diagonal elements of the of the
Green's function matrix, is simply given by
\begin{equation}
\textrm{Im}(G_{ee}) = \frac{J(e_x, e_t)}{|E_0|^2},\qquad \textrm{Im}(G_{tt}) = \frac{J(t_x, t_t)}{|T_0|^2},
\end{equation}
where $J(e_x,e_t)$ denotes that the current is evaluated on the first pair of solution $(e_x,e_t)$, and
similarly for $J(t_x,t_t)$. Note that the counter terms, which are explicitly real, do not alter the value of
the imaginary parts of the Green's function. The imaginary parts of the diagonal elements are thus proportional
to the conserved current, which from (\ref{horizoncurrent}), is shown to be positive definite, as is required by
definition. The off-diagonal elements are not proportional to the current because each of the field that appears
in the quadratic current is evaluated on a different solution.

The conserved current also serves as an important check of our numerics.



\subsubsection{Numerical solutions and dispersion relations}
We could identify numerically three distinct quasi-normal modes in the mixed system.

\FIGURE[!ht]{
\begin{tabular}[ht]{cc}
\tiny{(a)} & \tiny{(b)}\\
\includegraphics[width=0.45\textwidth]{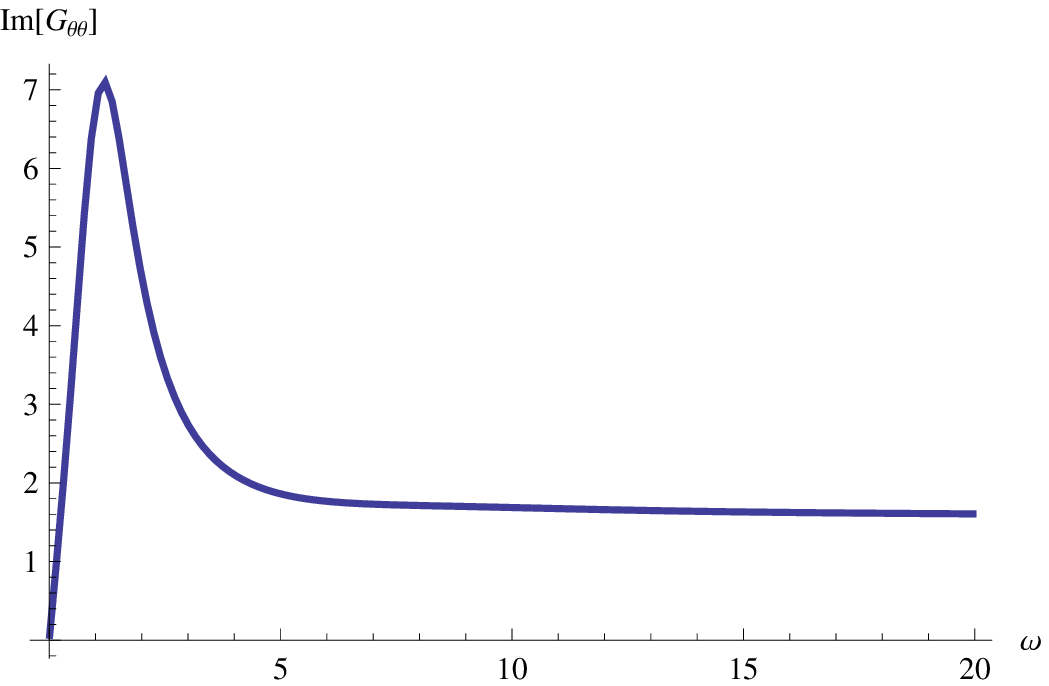}& \includegraphics[width=0.45\textwidth]{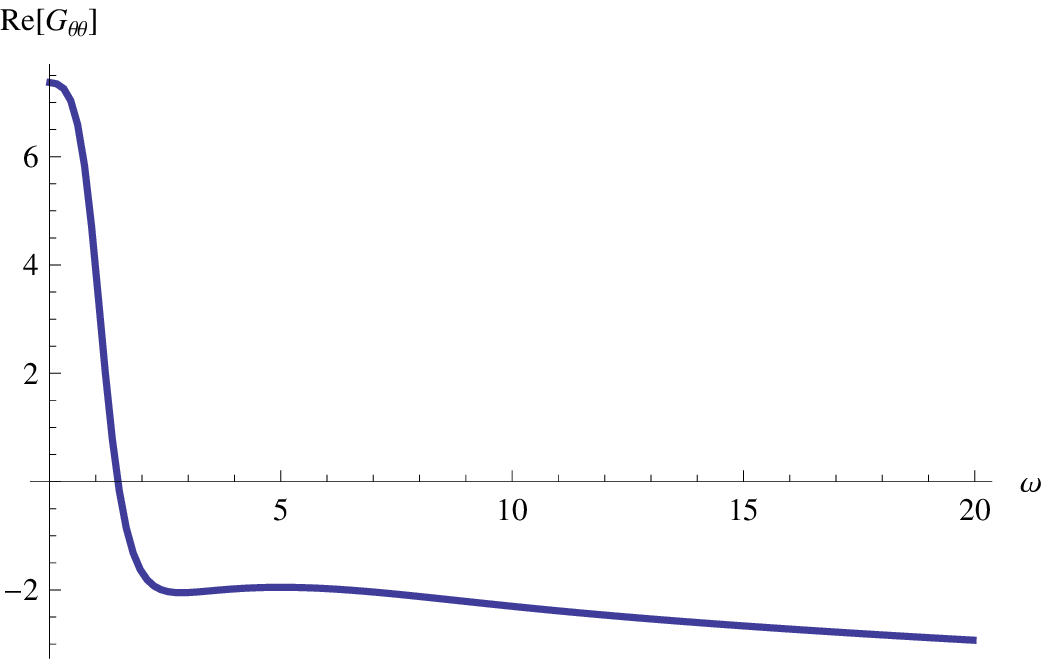}\\
\tiny{(c)} & \tiny{(d)}\\
\includegraphics[width=0.45\textwidth]{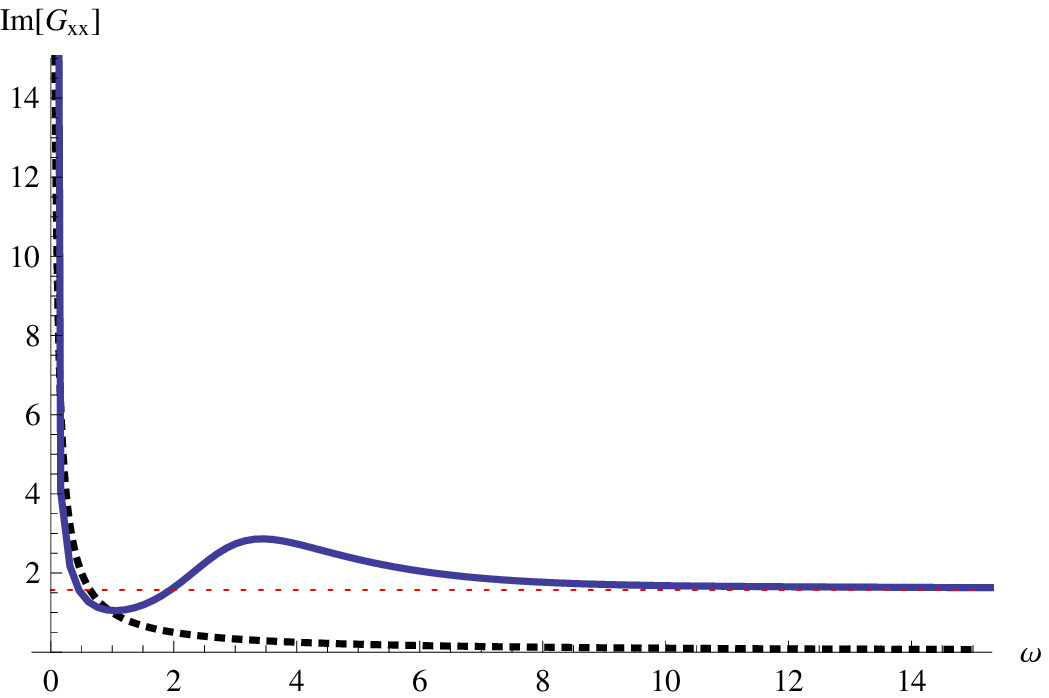}& \includegraphics[width=0.45\textwidth]{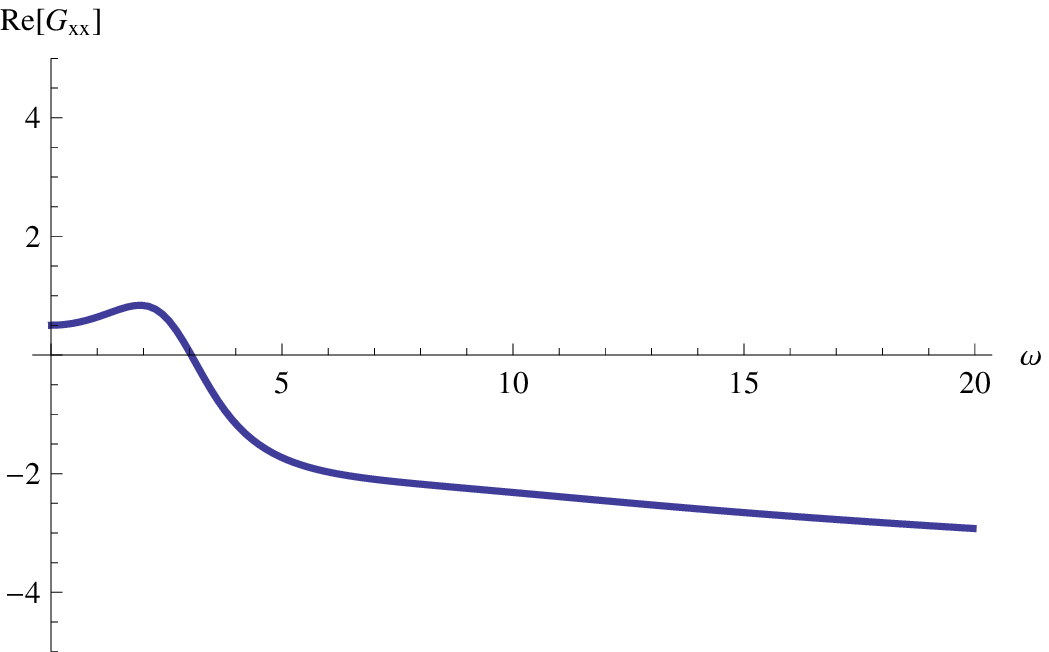}\\
\end{tabular}
\caption{The real and imaginary parts of the diagonal components of the Green's functions at $d/\bar{C}=1.074, k=0$. The dotted black curve
corresponds to $1/\omega$ and the dotted red horizontal curve corresponds to $\pi/2$. (All plots are obtained at $r_H=1$.)} \label{Greenfns}}
Figure (\ref{Greenfns}) presents
typical plots of the diagonal components of the Green's functions at vanishing $k$. Note that at
zero $k$ the equations of motion are decoupled and $t_x$ and $e_t$ are identically zero. As a result the
off-diagonal terms in the Green's function matrix is also zero.

The first peak we found corresponds to the divergence in
$\textrm{Im}(G_{xx})$  as $\omega \to 0$. At $k=0$ the
equations of motion decouple and the divergence in $\textrm{Im}(G_{xx})$ corresponds precisely to the dissipationless
quasi-normal mode as computed in the previous section in the hydrodynamic limit.
As already mentioned in section (\ref{soundmodes}) however, this mode remains there with unaltered dispersion even in the
finite $d$ and $q$ limit, where mixing becomes important.


The second quasi-normal mode we found correspond to the second peak of $\textrm{Im}(G_{xx})$ in (\ref{Greenfns}).
Typical plots of it's dispersion are given in figure (\ref{dispmixed}). The real part of the quasi-normal frequency begins roughly
quadratic in $k$, but then asymptotically approaches $\omega_R = k$. The imaginary part $\omega_I$ decreases
with increasing $q$, but then stabilises as $k$ continues to increase, and show no sign of instabilities as far
as we have checked.

\FIGURE[!ht]{
\begin{tabular}[ht]{cc}
\tiny{(a)} & \tiny{(b)}\\
\includegraphics[width=0.45\textwidth]{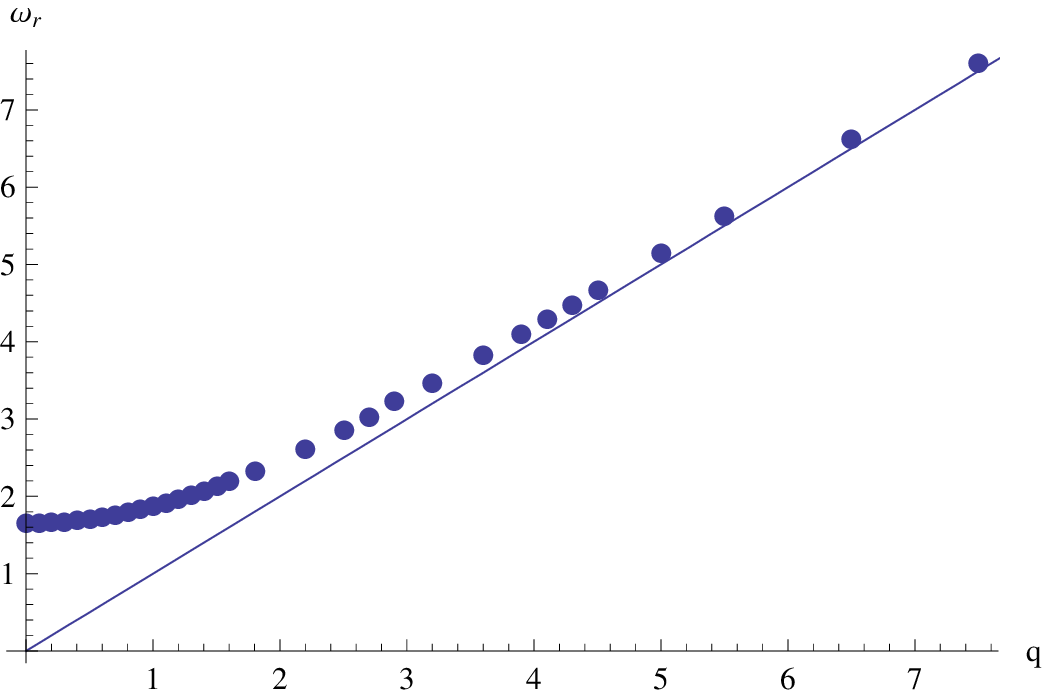}& \includegraphics[width=0.45\textwidth]{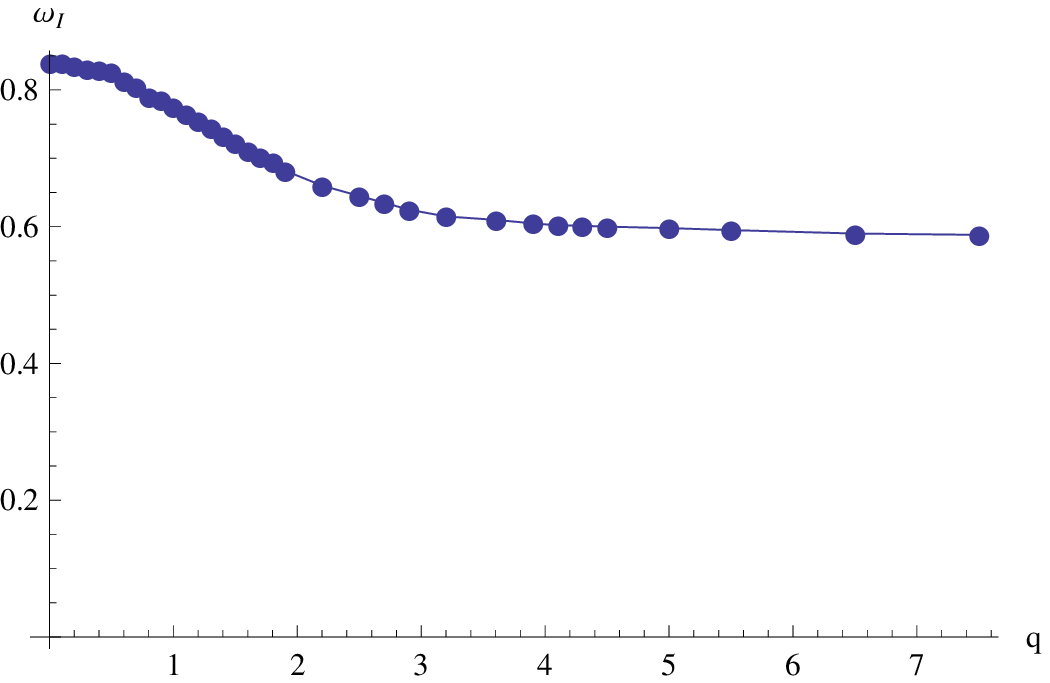}
\end{tabular}
\caption{(a) and (b) shows a typical dispersion relation of the quasi-normal mode corresponding to the second peak in $\textrm{Im}G_{xx}$.
Here $\omega = \omega_R - i \omega_I$. Similar behaviour was found in\cite{Myers:2008cj}.
} \label{dispmixed}}

The first peak in
$\textrm{Im}(G_{\Theta\Theta})$ in fact corresponds to a pole at $\omega = 0-i (a+ D k^2)$, for some constant
$a$ and $D$ which depends on $d$ and the embedding $\chi$. The peak gets shifted from $\omega=0$
because it happens that $t_1$ approaches zero also at  $\omega=0$. This is the third quasi-normal mode
we followed.
A typical plot of the dispersion of this pole is given in
figure (\ref{wIqchi}).

\FIGURE[ht]{
\includegraphics[width=0.6 \textwidth]{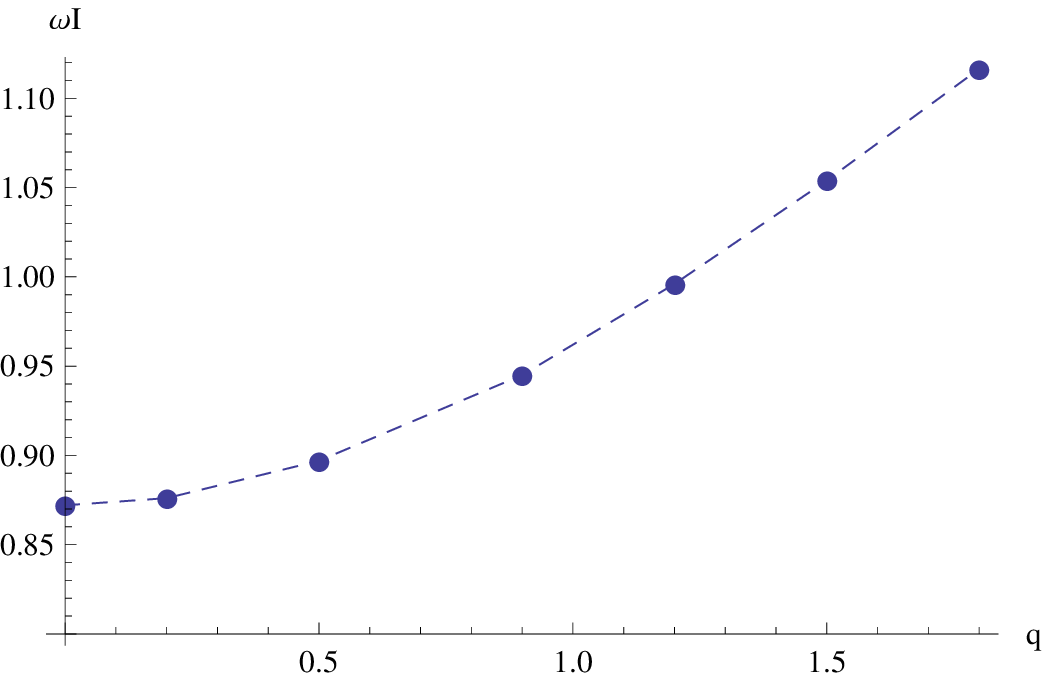}
\caption{Typical plot of the dependence of $\omega_I$ on momentum $q$ of the quasi-normal mode corresponding to the
 first peak in $\textrm{G}_{\Theta\Theta}$. } \label{wIqchi}}

Beyond these three peaks visible in the plots of the spectral functions, they look completely smooth and we
have not attempted to identify more quasi-normal modes.



Typical plots of the off-diagonal Green's functions at finite $k$ are presented in figure (\ref{Greenfnsqoff}).
It is not surprising that the imaginary part of $G_{x\theta}$ goes negative since space ($\mathcal{P}$) and time ($\mathcal{T}$) reversal symmetries
are explicitly broken by the background electric field. This is consistent with the observation in \cite{Hartnoll:2009sz} that
positivity of $\textrm{Im} G_{x\theta}$ is guaranteed only when $\mathcal{P}$ and $\mathcal{T}$ are individually preserved.
The diagonal components of the Green's functions at finite $k$ look very similar to the $k=0$ case, except that the peaks are shifted, and
therefore are not shown here. It is however important to point out that the imaginary parts of the diagonal elements remain positive definite
at all frequencies. From the bulk perspective this is guaranteed by the conserved current, as already explained in (\ref{conservedj}).
This is also required from the definition of the Green's function $G_{\mathcal{O}_A\mathcal{O}_A}(\omega)$ for any operator $\mathcal{O}_A$ in the dual field theory. The imaginary part of the Green's function is given
schematically by
\begin{equation}
\textrm{Im} G_{\mathcal{O}_A\mathcal{O}_A}(\omega) \sim \mathcal{Z}^{-1}\sum_{\alpha,\beta} |<\alpha|\mathcal{O}_A|\beta >|^2 (e^{-\beta \Xi_\alpha} \pm e^{-\beta \Xi_\beta}) \delta(\omega + \Xi_{\alpha\beta}),
\end{equation}
where $\Xi_\alpha= E_\alpha-\mu N_\alpha$, $\Xi_{\alpha\beta}= \Xi_\alpha-\Xi_\beta$, $\mathcal{Z}$ is the partition function, and the sum is over all eigenstates of the Hamiltonian.
The $+$ sign is taken for fermions and $-$ for bosons. It is clear that every term in the sum is positive.


\FIGURE[!ht]{
\begin{tabular}[ht]{cc}
\tiny{(a)} & \tiny{(b)}\\
\includegraphics[width=0.45\textwidth]{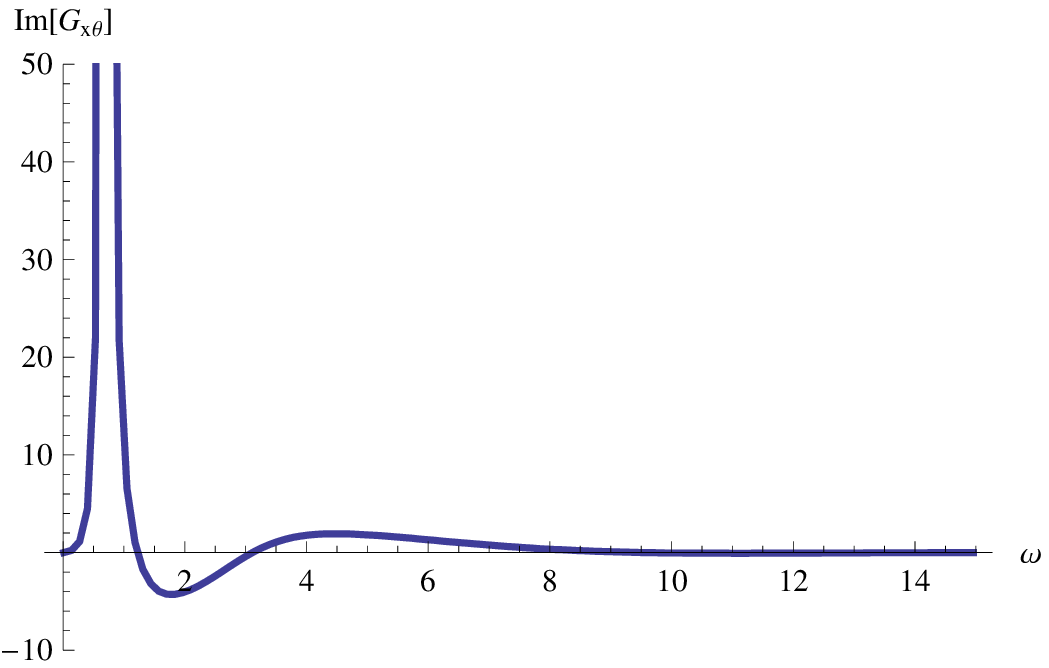}& \includegraphics[width=0.45\textwidth]{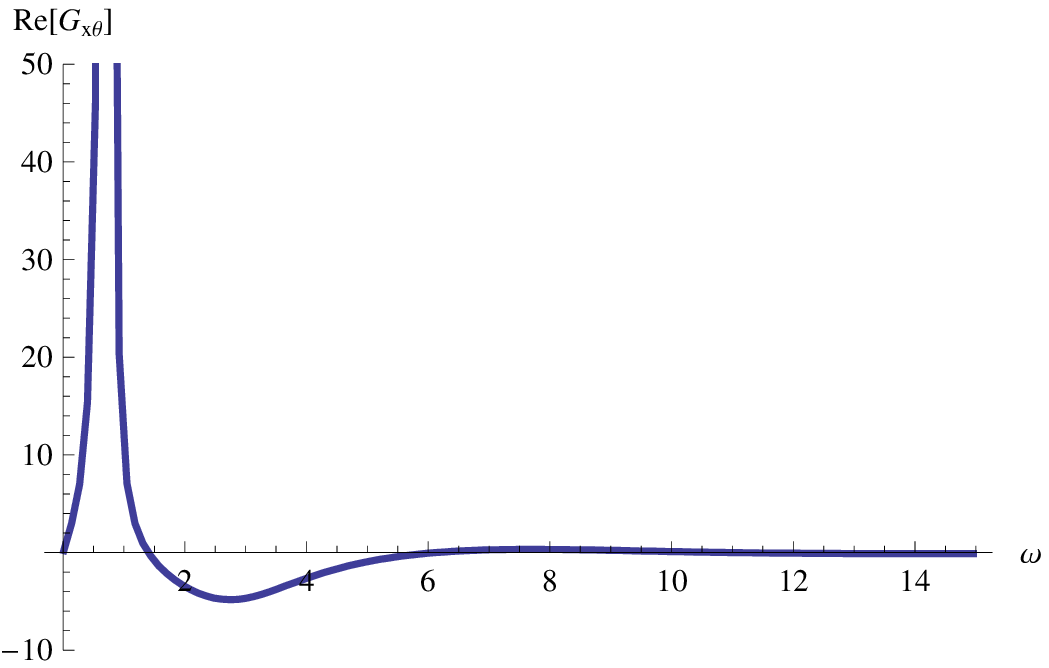}\\
\end{tabular}
\caption{The real and imaginary parts of the $G_{x\theta}$ at $d/\bar{C}= 1.074, k=0.7$ and $r_H=1$.} \label{Greenfnsqoff}}


We note that in the high frequency limit $\omega \gg 1$, the Green's function approaches that of the conformal result i.e. the vanishing
baryon density $d$ and temperature limit. The numerical result matches precisely the analytic result obtained in equation (\ref{gxx0td}).




\section{Fermionic fluctuations and spectral functions}

One of the most interesting aspects of a 1+1 dimensional system is the possibility of having spin-charge
separation. Roughly speaking, this corresponds to charged fundamental fermions behaving as bound states of two
degrees of freedom, one of which carries only spin and is called the spinon and the other which carries only
charge and is called a holon. We use this as a motivation for studying the spectral function of world volume
fermions. We however find no evidence of spin charge separation in these spectral functions.

The analysis follows \cite{LMV}. Firstly, in this setup, since we are considering a single D3 probe,
the fermions do not carry any U(1) charge. Hence the Dirac equation for fermions simply reads\cite{Aganagic:1996pe,Aganagic:1996nn}, \be
(G+F)^{\mu\nu}\Gamma_\mu D_\nu \psi=0\,, \ee where \be D_\nu=\partial_\nu+\frac{1}{4}\omega_\nu^{ab}\Gamma_{a
b}\,. \ee Here Greek letters will indicate curved space indices while Latin indices will represent tangent space
indices.

For simplicity we will only consider the $\chi=0$ case leaving the massive case for future work. Following \cite{LMV}, the fermion spectral function exhibits two independent components given by $\rm{Im} \xi_+$
and $\rm{Im} \xi_-$ where $\xi_\pm$ satisfy the following equations,

\be \xi_\pm'=\pm 2 D \xi_\pm +i (A \pm C)
+i (A \pm C) \xi_\pm^2\,, \ee
where
\be D=i \frac{F_{rt} k}{\sqrt{ g_{xx} |g_{tt}|}}\,,\quad C=i
\sqrt{\frac{g_{rr}}{g_{xx}}}k\,,\quad A=i\sqrt{\frac{g_{rr}}{|g_{tt}|}}+F_{rt}\frac{\partial_r g_{xx}}{4 g_{xx}
\sqrt{g_{rr}|g_{tt}|}}\,, \ee
with
\be F_{rt}=\frac{d}{\sqrt{d^2+r^2}}\,. \ee
Here we have assumed a Fourier
decomposition of the fermion fields with $w,k$ denoting the energy and spatial momentum respectively. Infalling
boundary conditions imposes
\be \xi_{\pm}=i\,, \ee \
at the horizon. It is easy to numerically solve for
the spectral functions. A typical plot is shown in figure \ref{ferm} (a). \FIGURE[ht]{\begin{tabular}[ht]{cc} \tiny{(a)} & \tiny{(b)}\\
\includegraphics[width=0.45 \textwidth]{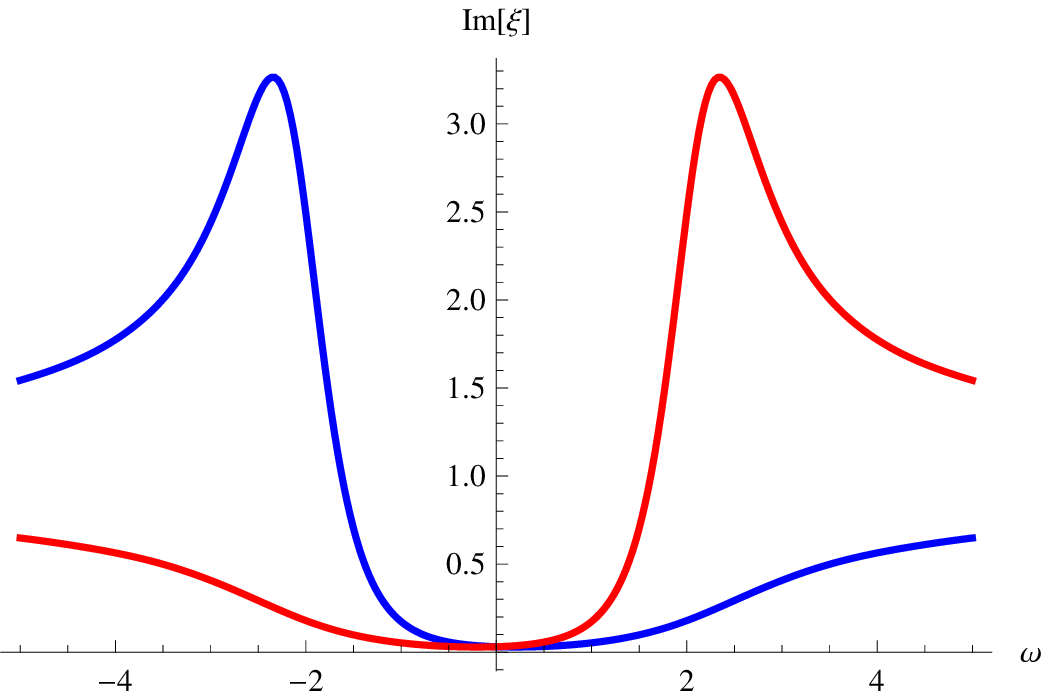}& \includegraphics[width=0.45 \textwidth]{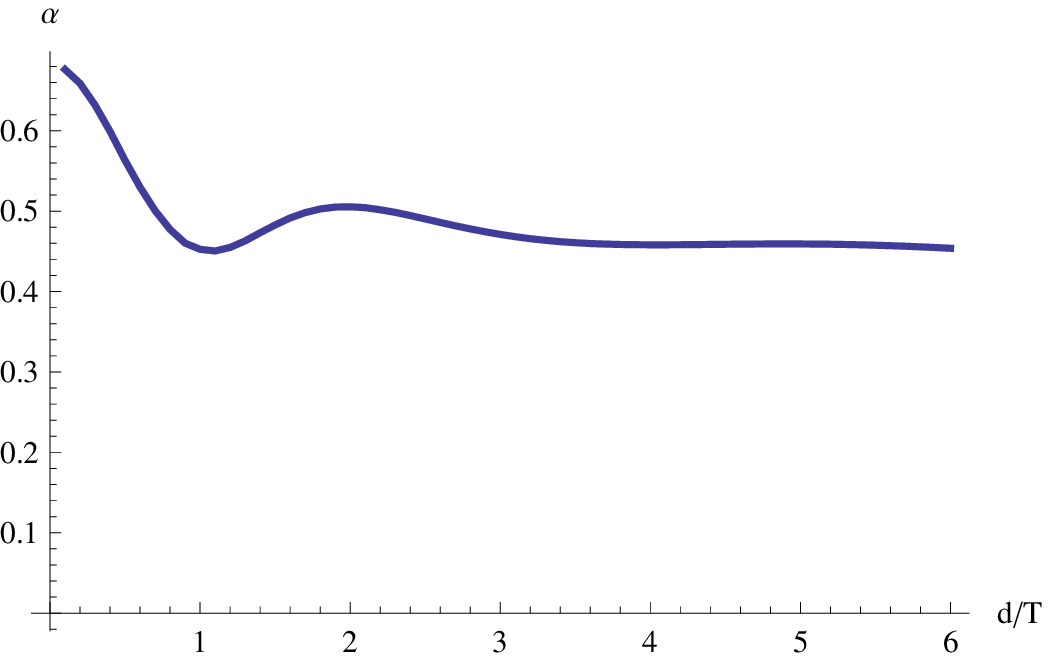}\\ \end{tabular}
\caption{(a) A typical plot of the fermion spectral functions. The right branch in red is $\xi_+$ while the left
branch in blue is $\xi_-$. (b) $\alpha$ vs $d/T$ for $\chi=0$, $r_H=1$.} \label{ferm}} We should point out that there is no particle-hole asymmetry due to
the absence of U(1) charge and also no sign of a fermi surface or a special momentum above which the
quasiparticles disappear. The dispersion relation for large $k$ is $\omega=\pm k$ with decreasing widths. If we do a fit for the height of the peaks as a function of $k$ of the form ${\rm Im} G(\Omega,k)= A k^\alpha$, we find a trend as in figure \ref{ferm}(b) suggesting a significant deviation from the Landau Fermi liquid picture where $\alpha=-1$ \cite{LMV}. Curiously for large $d/T$, $\alpha$ seems to asymptote to $\alpha_*\approx 0.45$. For low $d/T$, $\alpha$ is around $0.68$. The exponent $\alpha/2$ is the ``anomalous dimension" of the fermion operators \cite{Voit}. From this we can extract the parameter $K$ for the Luttinger model which is given by
\be
\alpha=\frac{1}{4}(K+\frac{1}{K}-2)\,.
\ee
Using this and assuming a repulsive interaction in the effective Luttinger model (in effect choosing the root $K<1$), we find $K$ lies between $0.22-0.28$.
It is very interesting to compare the value for $\alpha$ for a massless fermion in the BTZ background, which can be extracted analytically. Considering the case that $T_L= T_R$ and making use of the analytic Green's function of a fermion in this background\cite{Iqbal:2009fd}, and setting the fermion mass $m=0$, we found that in the large $k$ limit the height of a QNM peak scales as $k^{h_R - \tilde{h}_R}= k^{1/2}$, where $h_R= 3/4, \tilde{h}_R= 1/4$ as defined in \cite{Iqbal:2009fd}. This gives $\alpha = 1/2$ and $K$ evaluates to $K= 2-\sqrt{3}\approx 0.27$, which is very close to what we found above numerically in the large $d$ limit.
{\it Curiously, the value $K=0.28$ is used to describe a wide range of experimental results in carbon nanotubes\cite{bockrath}}!

\section{Conductivities}
The conductivity can be extracted from the Green's function. At zero momentum, the conductivity is related
to the current-current correlation function simply by
\begin{equation}
\sigma(w) = -\frac{i G_{J_xJ_x}(\omega)}{w},
\end{equation}
whose temperature dependence at constant baryon density $d$ can be readily obtained. Note that here we denote the \emph{dimensionful} frequency by $w$, which is defined as $w= r_H\omega$\footnote{The distinction between them is only important in this section where we would like to vary the temperature, instead of setting $r_H=1$.}.  The results are shown in
figure (\ref{conductplotbare})\footnote{Note that we have not included the logorithmic dependence $\log r_H$ in the imaginary part of the conductivity. }.


\FIGURE[h]{
\begin{tabular}[h]{cc}
\tiny{(a)} & \tiny{(b)}\\
\includegraphics[width=0.45\textwidth]{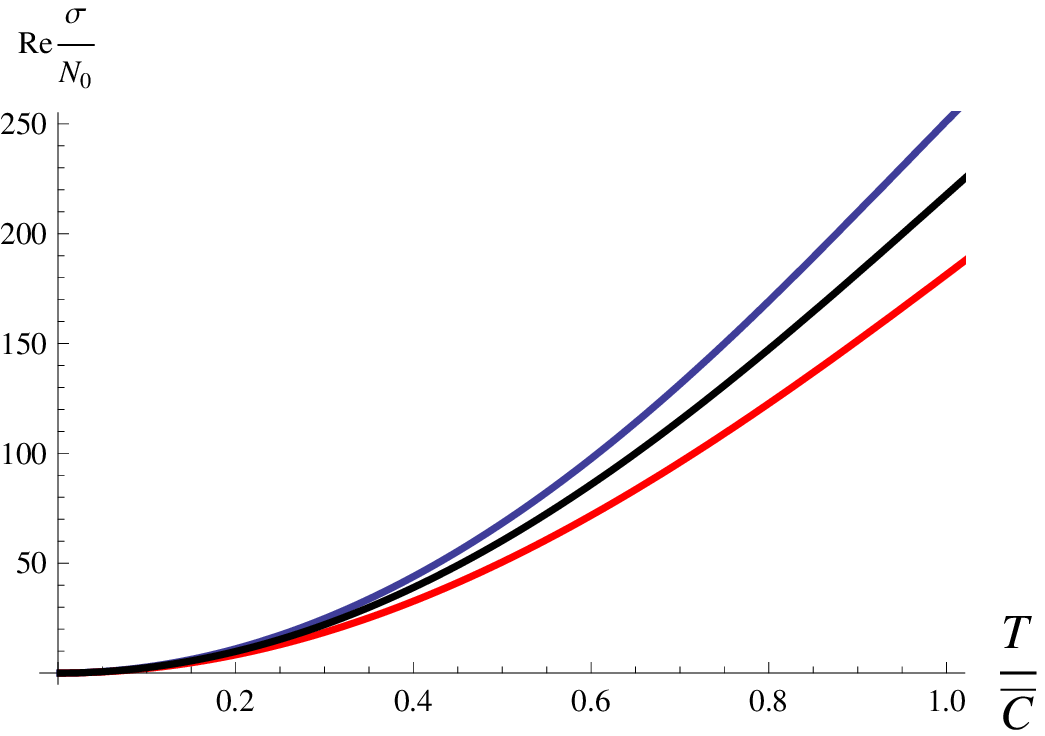}& \includegraphics[width=0.45\textwidth]{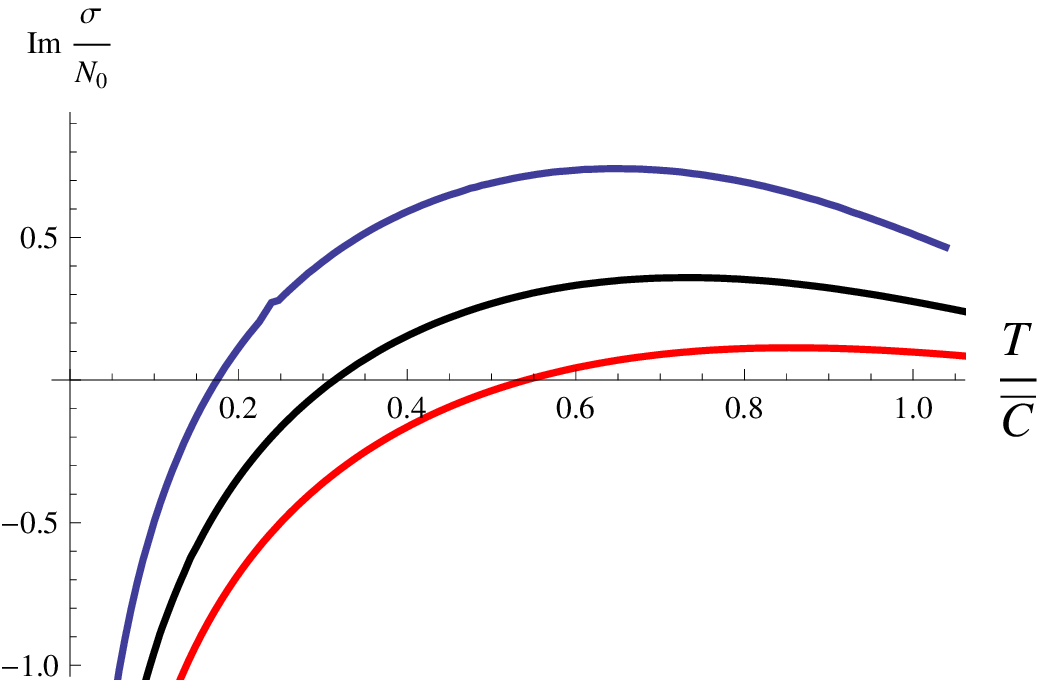}\\
\end{tabular}
\caption{(a) and (b) show plots of the real and imaginary parts respectively of the conductivities as a function
of temperature at  $q=0$ and $\omega = w/r_H =0.01/(\pi T)$, where $d/\bar{C}=1.04 ~\textrm{(blue curve)}, 1.10 ~\textrm{(black curve), and~}
1.20 ~\textrm{(red curve)}$.} \label{conductplotbare}}
From figure (\ref{conductplotbare}), it is clear that the real part of the conductivity,
becomes independent of $T$ in the high temperature limit.
Re $(\sigma)$ behaves like
\begin{equation}
\textrm{Re}(\sigma(w/d, w/\overline{C}\ll 1))|_{{\rm small}~T} \sim T^0, \qquad \textrm{Re}(\sigma(w/d, w/\overline{C}\ll 1))|_{{\rm large}~T} \sim T,
\end{equation}
 This can be compared with a normal Fermi liquid, whose real dc conductivity falls off as $1/T^2$ in the low temperature limit. It is also
interesting to compare this with a simple version of a spin half Luttinger liquid (see for example \cite{Voit}),
whose low frequency and low temperature conductivity can be parametrised in terms of the interaction strength,
\begin{equation}
\textrm{Re}( \sigma_{\textrm{dc}}) \sim T^{3-p^2 K_\rho},
\end{equation}
where $1/p$ is the band filling, and $K_\rho$ characterises the interaction strength of the
charge density waves\cite{Voit}.

Using the result for the Green's function in section (\ref{soundmodes},\ref{mixedspectral}), we obtain also, at
small frequencies,
\begin{equation}
\textrm{Re}(\sigma(w)) \sim w^{-2}, \qquad \textrm{Im}(\sigma(w)) \sim w^{-1}.
\end{equation}
Note that the $1/w$ behaviour of $\textrm{Im}\sigma(w)$ is deduced from the fact that $\textrm{Re}(G_{xx})$ approaches a constant as $\omega\to 0$.  This signals that the real part of the conductivity should contain a delta function $\delta(w)$\cite{Hartnoll:2009sz} on top of the $1/w^2$ divergence, although the delta function would not show up in the numerics or a perturbative expansion in $w$, since $w$ is never taken to be strictly zero.
We again compare with the Luttinger liquid behaviour given by
\begin{equation}
\textrm{Re} (\sigma(w)) \sim w^{p^2K_\rho-5}.
\end{equation}

It is tempting to fit our results with those of the simple Luttinger liquid discussed above. At larger values of the temperature, where the conductivity approaches a constant, our results for the real part of the conductivity can be roughly fitted with the Luttinger liquid, with the parameter $p^2K_\rho \approx 3$.  Here $K_\rho,p$ are treated as some effective parameters, a combined effect of all the degrees of freedom there in our theory.

Typical plots of the conductivity vs frequency at constant temperature is given in figure (\ref{conduct_w}). Note that the real and imaginary parts are diverging at different powers of $\omega$ as $\omega \to 0$.
\FIGURE[h]{
\begin{tabular}[h]{cc}
\tiny{(a)} & \tiny{(b)}\\
\includegraphics[width=0.45\textwidth]{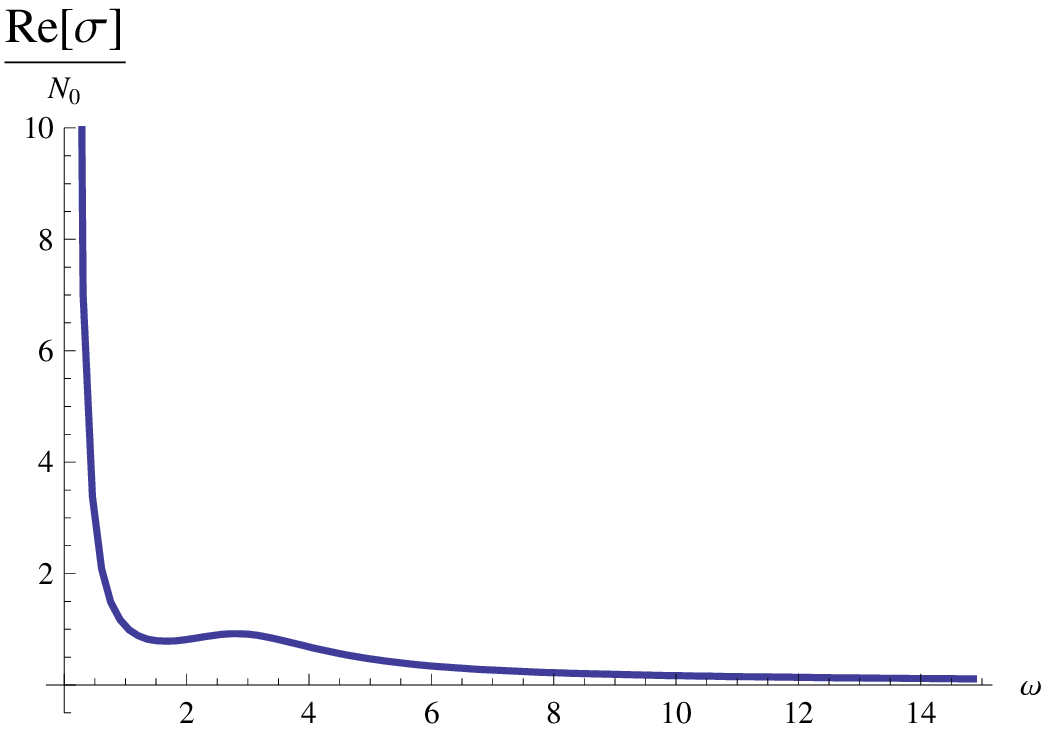}& \includegraphics[width=0.45\textwidth]{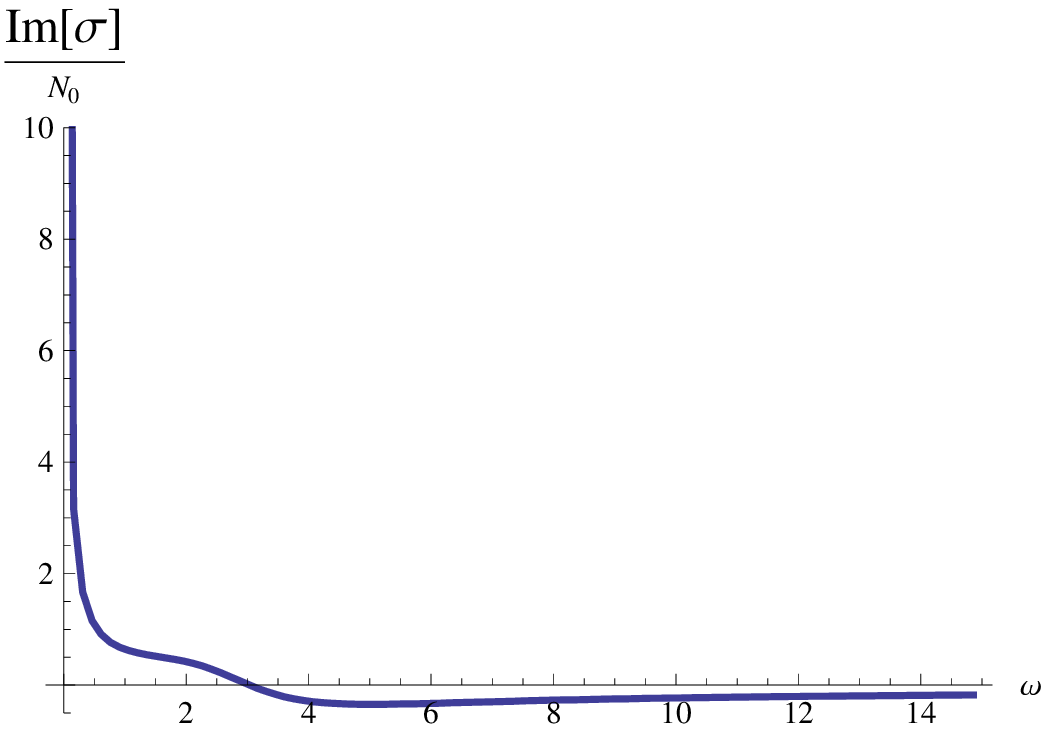}\\
\end{tabular}
\caption{(a) and (b) show typical plots of the real and imaginary parts respectively of the conductivities as a function
of the frequency at  $q=0$, $r_H=1$. } \label{conduct_w}}

\section{Discussion}

We studied 1+1 dimensional holographic quantum liquids using a D3-D3
defect CFT. We set up the thermodynamics
and explained how the boundary terms differ when compared to higher
dimensions. Unlike higher dimensions, it is
necessary to introduce a gauge potential counterterm to remove the log
divergence arising from it. We explained
how to compute both bosonic and fermionic spectral functions in this
setup. Unlike higher dimensions we found a
{\it dissipationless} mode and a {\it purely dissipative} mode. The dissipationless mode is a pole in the density fluctuations, and can thus be interpreted as a sound mode. This is what is generally found in a Luttinger model after diagonalisation by bosonisation, from which one obtains the charge density wave as an eigen-fluctuation of the Hamiltonian, which disperses at a particular velocity $v$ depending on the coupling of the mother fermions\cite{Haldane1,Voit}. In our case, the velocity is simply $v=1$, taking the conformal value\footnote{This is probably a result of working in the probe limit, where we are breaking conformal invariance ``weakly". Deviation from the conformal result is probable upon taking into account back reaction.}.

To the best of our knowledge, the latter purely dissipative mode
is not present in any Luttinger
model. It could be that the appearance of this mode is a large $N$ artifact. This issue clearly merits further study. In the extended Hubbard model in 3+1 dimensions \cite{mahan}, a purely dissipative spin mode and a non-dissipative charge mode was found. We are not aware of similar studies for the extended Hubbard models in 1+1 D.  In our case, when $d=0$, the dissipative mode disappears and it is likely that a Luttinger description holds and one may be able to extract the parameters $v$ and $K$ used in this description. We find that $v=1$ in all cases. For the fermions, assuming an effective Luttinger description, it was found that the parameter $K>0.22$. In fact for a wide range of $d/T$ we found $K\approx 0.28$. This value is reported in the literature \cite{bockrath} as describing a wide range of phenomena in carbon nanotubes. A simple estimate of $K$ from the Luttinger model, making explicit use of the graphite band structure gives $K \sim 0.2$\cite{PhysRevLett795086}, which demonstrates that $K$ is a material dependent quantity. In fact more recently studies of Gallium Arsenide quantum wires gives a value of $K\sim 0.66$\cite{2006PhRvL97s6802L}. These results, combining with the plot of $\alpha$ vs $d/T$ in figure (\ref{ferm}) would suggest that there is an upper bound on $\alpha$, and thus a lower bound{\footnote{One of the achievements of AdS/CFT is to come up with the famous viscosity bound $\eta/s>1/4\pi$ \cite{etabys,etabys1,Kovtun:2004de}. Although the precise number is doubtful \cite{etabys2,etabys3}, it is thought that a bound exists. In a similar spirit, one can look for other physical quantities where AdS/CFT methods would lead to a bound.}} on $K>0.22$. Clearly this deserves a closer look. It should be possible to determine the effective $K$ and $v$ for the bosonic sectors in the probe brane model but for this one needs to work out the fermionisation map between the Luttinger fermions and the AdS bosons. Moving the branes away from the origin of the moduli space \cite{erd} may also lead to different physics. A model that will be
interesting to study is the D3-D7 setup
making a 1+1 dCFT. This model will have chiral fermions. It would be interesting to compare with what is known about chiral Luttinger models\cite{chiralLuttinger}. A first step in this direction has been taken in \cite{Fujita:2009kw}.
Moreover, it will be interesting to answer: What is the holographic analogue of spin charge separation? In the spin-half Luttinger liquid, the charge and spin degrees of freedom are identified with $n(k)_\uparrow \pm n(k)_\downarrow$ respectively, where $n(k)_{\uparrow,\downarrow}$ correspond to the Fourier components of the density operator of the spin up and down fermions respectively\cite{Voit}. We have studied in this paper the Green's function of the charge density of the world-volume $U(1)$, which we can naturally identify as the Green's function of the charge density wave, whose dual operators involve the  bilinears of fermions belonging to the hypermultiplets\cite{erd}. We will treat the R-symmetries indices of the fermions in the dual dCFT as spin labels. One could then look for suitable fluctuations on the probe brane which would allow us to compute the Green's function of the spin-density wave. A candidate of such a fluctuation is that of $A_{\zeta_1}$ and one could generally expect that the quasi-normal modes would disperse differently from the charge density degree of freedom\cite{progress}, leading to a natural realisation of spin-charge separation. So far we have identified the lowest mode of $A_{\zeta_1}$ which is also a purely dissipative mode. This is already suggestive of a spin-charge separation, with the spin density wave dispersing at $v=0$ and the charge density wave at $v=1$\cite{progress}.

One  feature that exists in Luttinger liquids is the Schottky anomaly in the specific heat \cite{suga-2008}. This happens when the number of states available in the microcanonical ensemble is finite and results in a bump in the specific heat. In Luttinger liquids, this is expected to happen at low temperatures. Although no such bump happens for the $\chi=0$ case, we have not been able to convince ourselves about the accuracy in our numerics to rule this out at $\chi\neq 0$. Moreover, as shown in appendix A, for small $\chi, \tilde{d}$ one can have an analytic handle on the specific heat. We find there that for non-zero $\chi$, the difference in the specific heat from the zero $\chi$ result is proportional to $c^2$ and is positive. Since for large temperatures, both cases will lead to $c_v\sim \pi T$, it seems that one prerequisite to have a bump is that the change in the specific heat from the $\chi=0$ case is positive! So it could still be that a better handle on the numerics will lead to something interesting like the Schottky type specific heat.
\FIGURE[ht]
{\includegraphics[width=0.4 \textwidth]{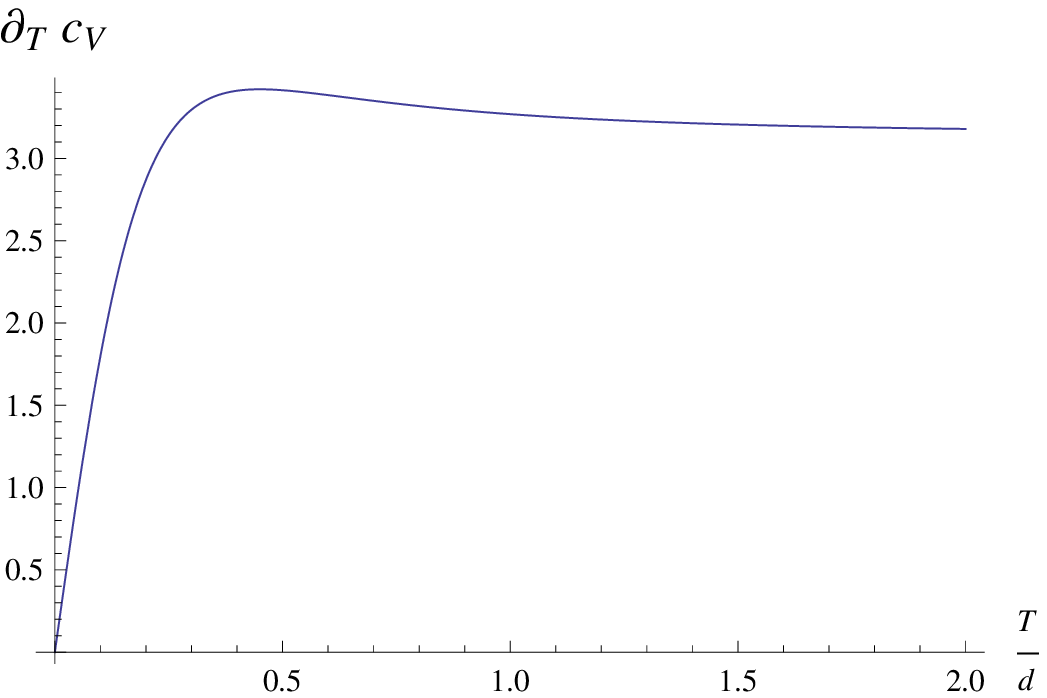}
\caption{$\partial_T c_v$ vs $T/d$ for $\chi=0$. The maximum occurs at $T=\sqrt{2}d/\pi$. } \label{diss1}}

We would like to conclude with the following observation. Since the low temperature specific heat behaves like $c_v \propto T^2$ and the high temperature specific heat behaves like $c_v\propto T$, it is clear that in the intermediate regime, $\partial_T c_v$ should have an interesting feature. As shown in figure (\ref{diss1}), we indeed find that for $\chi=0$, $\partial_T c_v$ exhibits a maximum for $T=\sqrt{2}d/\pi$. As we turn on $\chi$ we expect the position of this maximum to get shifted to the left (since generically the combination $\sqrt{\tilde{d}^2-c^2}$ appears in place of $\tilde{d}$) and eventually vanish near $\tilde{d}\sim c$. The fact that the low temperature specific heat behaves like $T^\ell$ and the high temperature one behaves like $T^h$ with $h<\ell$ appears quite generic in these models. Hence we expect this maximum in $\partial_T c_v$ to persist in all such defect CFTs.

After this work was completed \cite{Maity:2009zz} appeared which deals with related issues.

\acknowledgments
It is a pleasure to thank Alex Buchel, Sean Hartnoll, Jaume Gomis, Bob McNees, Rob Myers, Miguel Paulos, Andrei Starinets, Javier Tarrio and David Tong for useful discussions. We thank Diptiman Sen for pointing out the reference to $K=0.28$ and alerting us to this result. AS thanks Greg van Anders for collaboration on the fermion Green function in related contexts. Research at Perimeter Institute is supported by the Government of Canada through Industry Canada and by the Province of Ontario through the Ministry of Research \& Innovation. A significant part of the work was completed while L-Y. H was a student at DAMTP. She would like to thank the Gates Trust for financial support.

\appendix

\section{Thermodynamics at small $\tilde{d}$ and $\chi$}
We have discussed the thermodynamics of the system at both vanishing and finite $\chi$. We find that at finite $\chi$ numerical solutions of
the embedding are used in the computation. We can in fact extract further analytic result in the limit of small $\tilde{d}$ and $\chi$, and expand the bulk action up to quadratic order in $\tilde{d}$ and $\chi$.

Doing that, we have
\be
E_r = \frac{d}{r}, \qquad A(r) = d \log \frac{r}{r_H},
\ee
which means $\mu = \overline{\mu}+ d\log r_H =0 $.

The solution for $\chi$ is given by
\be
\chi(u)= \frac{\overline{C} \sqrt{\pi} [\Gamma(1/4)^2 {_2F_1}(1/4, 1/4, 1/2, u^4) -
   2 u^2 \Gamma(3/4)^2 {_2F_1}(3/4, 3/4, 3/2,
     u^4)]}{4 r_H \Gamma(1/4) \Gamma(3/4)},
\ee
where we have fixed the boundary condition at the horizon such that it is regular, and that at the AdS boundary the
coefficient of the leading $\log$ term is simply $c=\overline{C}/r_H$.

We can substitute these expressions into the action and the counter terms. This gives
\begin{eqnarray}
\frac{\mathcal{F}}{N_0}&=& r_H^2(-\frac{1}{2} - \frac{c^2 \log 8}{4}- \frac{(\tilde{d}^2-c^2)}{2}\log r_H),\nonumber \\
\frac{S}{N_0} &=& r_H(1+ \frac{(\tilde{d}^2-c^2)}{2}),\qquad \frac{c_v}{N_0}= r_H(1- \frac{(\tilde{d}^2-c^2)}{2}),
\end{eqnarray}
which agrees with the small $d$ expansion of the analytic result (\ref{sant},\ref{cvant}) at zero $c$.

The expression $(\tilde{d}^2-c^2)$  is strictly positive in the stable region and negative in the unstable region.

\section{The mixed modes Lagrangian}

We would like to specify in the table below the leading near horizon and boundary behaviour of the various functions we defined in equation (\ref{mixedlang}).

\begin{equation}
\begin{array}{|c|c|c|}
\hline
    &   u\to 1& u\to \infty \\
\hline
&&\\
T_1 & -\frac{1 - \chi_0^2}{ 2 \sqrt{1 + \tilde{d}^2 - \chi_0^2} }         & -\frac{u}{2}\\
&&\\
\hline
&&\\
T_2 & \frac{1 - \chi_0^2}{2 \sqrt{1 + \tilde{d}^2 - \chi_0^2}}& \frac{1}{2u}\\
&&\\
\hline
&&\\
T_3 &-\frac{\sqrt{1 + \tilde{d}^2 - \chi_0^2}}{8 (u - 1)} & -\frac{1}{2u}\\
&&\\
\hline
&&\\
T_4 & 2 \sqrt{1 + \tilde{d}^2 - \chi_0^2}(u - 1)& \frac{u^3}{2}\\
&&\\
\hline
&&\\
T_5 &\frac{\chi_0^2 (u - 1)}{\sqrt{1 + \tilde{d}^2 - \chi_0^2}}  & c^2 (\log u)^2\\
&&\\
\hline
\hline
&&\\
D_1&-\frac{\tilde{d} \chi_0 \sqrt{1 - \chi_0^2}}{4(1 + \tilde{d}^2 -  \chi_0^2)} & -\frac{c \tilde{d} \log u}{u^4}\\
&&\\
\hline
&&\\
D_2&\frac{\tilde{d} \chi_0 (u - 1)}{\sqrt{1 - \chi_0^2}} & c \tilde{d}\log u\\
&&\\
\hline
&&\\
D_3&-\frac{\tilde{d} \chi_0}{\sqrt{1 - \chi_0^2}} & -\frac{c\tilde{d}\log u}{u}\\
&&\\
\hline
\hline
&&\\
S_1 &-\frac{\sqrt{1 + \tilde{d}^2 - \chi_0^2}}{8 (u - 1)}& -\frac{1}{u^3}\\
&&\\
\hline
&&\\
S_2 &-\frac{(1 + \tilde{d}^2 - \chi_0^2)^{3/2}}{2 (1 - \chi_0^2)}&-\frac{u}{2}\\
&&\\
\hline
&&\\
S_3 &2 \sqrt{1 + \tilde{d}^2 - \chi_0^2} (u - 1)&\frac{u}{2}\\
&&\\
\hline
\end{array}
\end{equation}

\section{More on stability}
Singularities of the spectral functions generally occur in the complex $\omega$ plane. When $\omega$ is complex,
$\Theta_{-k}$ and similarly $a_{\mu,-k}$ cease to be the complex conjugates of $\Theta_k$ and $a_{\mu,k}$
respectively. We can, however, still obtain some analytic results on the stability of the system (i.e. determine
the sign of $\omega_I$ at which the singularity is located). To that end, let us take equation
(\ref{ateom})multiplied by $\bar{a}_{t,k}$, and subtract it by $a_{x,k}$ multiplied by the complex conjugate of
(\ref{axeom}) and finally adding to them $\bar{\theta}_{k}$ times (\ref{thetaeom}). The resultant equation is
given by (we will omit $k$ in the expression)
\begin{eqnarray}
&& 2 \bar{a}_{t} \partial_u(S_2 a_t')-2a_x\partial_u(S_3 \bar{a}_x') + 2\bar{\Theta}\partial_u(T_4 \Theta')+(T_5'-2(T_1+k^2T_2+\omega^2T_3))|\Theta|^2 \nonumber\\
&&-2k^2S_1|a_t|^2 + 2\bar{\omega}^2 S_1|a_x|^2 + 2k \bar{a_t}a_x S_1(\bar{\omega}-\omega) \nonumber\\
&&+D_1 k(\bar{\omega}\bar{\Theta}a_x-k\bar{a}_t\Theta)-\bar{a}_t\partial_u(D_3\Theta + D_2\Theta')\nonumber\\
&&-D_1 k(k a_t \bar{\Theta}+\omega a_x\bar{\Theta})+ \bar{\Theta}(D_3 a_t'-\partial(D_2 a_t'))=0,
\end{eqnarray}
where the primes also denotes partial derivatives w.r.t. $u$. Taking the imaginary part and integrating by
parts, we are left with
\begin{eqnarray}\label{exact1}
&&\textrm{Im}[\partial(2 \bar{a}_t S_2 a_t' -2a_x S_3 \bar{a}_x' + 2\bar{\Theta}T_4 \Theta' + a_tD_3\bar{\Theta}+a_tD_2\bar{\Theta}' + \Theta D_2\bar{a}_t')]\nonumber\\
&& = -\omega_I (4\omega_R(T_3|\Theta|^2 + S_1|a_x|^2)+D_1 k (a_x\bar{\Theta}+\bar{a}_x\Theta)+
2kS_1(\bar{a}_ta_x + \bar{a}_xa_t)),
\end{eqnarray}
where $\omega=\omega_R - i \omega_I$. The cross terms on the r.h.s. of (\ref{exact1})can be replaced by a single
term using (\ref{axeom}). After some simplification we are finally left with
\begin{eqnarray}\label{identity1}
&&\partial_u \{\textrm{Im}(2 \bar{a}_t S_2 a_t' -2a_x S_3 \bar{a}_x' + 2\bar{\Theta}T_4 \Theta' )+ \frac{\omega_I}{|\omega|^2}\textrm{Re}(2\bar{\omega}\bar{a}_xS_3a_x')\}\nonumber \\
&& =-\omega_I (4 \omega_R T_3 |\Theta|^2 -2 S_3 |a_x'|^2 \frac{\omega_R}{|\omega|^2}).
\end{eqnarray}

We have dropped cross terms inside the partial derivatives since we will eventually integrate over $u$ from the
horizon to the boundary, and the $D$ terms vanish at both limits. The expression can be further simplified using
(\ref{axEx},\ref{atEx}), and explicitly integrating over $u$ from the horizon to the boundary, we have
\begin{eqnarray}\label{identity2}
&&\bigg\{ 2\textrm{Im}[S_3 \frac{k \bar{a}_t+ \omega_R\bar{a}_x}{\omega^2-k^2}E_x'+ \bar{\Theta}T_4 \Theta' ]\bigg\}\bigg\vert_{u\to\infty}+ 2\omega_R\sqrt{1+d^2-\chi^2_0}(u-1)^{-\omega_I/2}\bigg\vert_{u\to 1}\nonumber\\
&&= \omega_I \omega_R \int_{1}^\infty V(u),
\end{eqnarray}
where \be V(u)= 4(-T_3)|\Theta|^2 + 2S_3\frac{|a_x'|^2}{|\omega|^2}. \ee We have also made used of the explicit
form of the solutions of $E_x$ and $\Theta$ at the horizon. From the explicit form of $T_3$ which is negative
definite and $S_3$ which is positive definite, we can conclude that $V(u)$ is positive definite.

Since the quasi-normal modes appear at the same time in $\Theta$ and $E_x$, the first term evaluated at the
boundary is zero at a quasi-frequency. The l.h.s is evaluated explicitly at the horizon, and is positive
definite. Therefore it implies that $\omega_I$ has to be positive, at the level of our approximations, where
these perturbations are small. The analysis here does not guarantee stability, since the same analysis would
have been valid in the case of a D7 probe, where we are well aware of the presence of instability from
\cite{Kobayashi:2006sb,Myers:2007we,Mateos:2007vc}.
It is important to note that if the quasi-normal mode occurs at $\omega_R=0$, the argument breaks down and our
equations of motion could produce a quasi-frequency where $\omega_I$ takes a negative value. In fact we found
precisely such a mode which gives a negative $\omega_I$ where the heat capacity turns negative, indicating
instability.

\section{The explicit form of the Green's functions from the probe D3-brane}
We would like to work out explicitly the form of the Green's functions, including the counter terms for the
probe D3-branes along the lines outlined in section (\ref{mixedspectral}).

Consider the probe D3 in the AdS black hole background. Ignoring for the moment the correct normalization, let
us write the solutions in the boundary limit as
\begin{eqnarray}
e_x = e_l \log[u] + e_0 + \mathcal{O}(\frac{1}{u^2}), & e_t =\frac{e_{1}}{u} + \mathcal{O}(\frac{1}{u^2}),\nonumber \\
t_x = t_0 + \mathcal{O}(\frac{1}{u^2}), & t_t = t_l \frac{\log[u]}{u} + \frac{t_1}{u}, \nonumber \\
\end{eqnarray}
and recall that
\begin{displaymath}
\Theta = E_0 e_t + T_0 t_t, \qquad E_x = E_0 e_x + T_0 t_x
\end{displaymath}
For notational simplicity, we denote $\{\Theta_{-k},{E_x}_{-k}\} = \{\bar{\Theta}_{k},\overline{E_x}_{k}\}$.
This relation however, holds literally only when $\omega$ is real. We have
\begin{equation}\label{bareGreen}
\left(\begin{array}{cc}
M_{\bar{e}e} & M_{\bar{e}t}\\
M_{\bar{t}e} & M_{\bar{t}t}
\end{array}\right) = \lim_{u \to \infty} \frac{-1}{2} \left(\begin{array}{cc}
2 \frac{S_2}{\omega^2-k^2} \bar{e}_xe_x'  & \frac{S_2}{\omega^2-k^2} \bar{e}_x t_x' + T_4 t_t \bar{e}_t'\\
\frac{S_2}{\omega^2-k^2} e_x \bar{t}_x'+ T_4 \bar{t}_te_t' & 2 T_4 \bar{t}_tt_t'
\end{array}\right),
\end{equation}
which explicitly evaluates to
\begin{eqnarray}
M_{\bar{e}e}&=& |e_l|^2 \log[u_\infty]+ \bar{e}_0e_l - |e_1|^2 , \nonumber \\
M_{\bar{e}t}&=& \frac{1}{2}\left(\bar{e}_1(t_l-t_1)-\bar{e}_1 t_l \log[u_\infty]-(\bar{t}_l e_1 \log[u_\infty]+\bar{t}_1e_1) + \bar{t}_0 e_l\right), \nonumber \\
M_{\bar{t}e}&=&  \overline{M_{\bar{e}t}},   \nonumber \\
M_{\bar{t}t}&=& -|t_l\log[u_\infty]+t_1|^2 +|t_l|^2 \log[u_\infty] + \bar{t}_1t_l,
\end{eqnarray}
so that the Green's function matrix again appears to be divergent. Here we have absorbed a factor of $1/(\omega^2-k^2)$ into $e_x$ and $t_x$ momentarily to avoid clumsy notation.

At the same time, counter terms have to be
taken into account. Let us therefore evaluate the counter terms in (\ref{count1},\ref{count2}), where the
quadratic terms are given by
\begin{equation}
\mathcal{C} = \int d^2x \left\{u_\infty^2 \bar{\Theta}\Theta (1-\frac{1}{\log[r_H u_\infty]})- \frac{A_\mu A_\nu
\eta^{\mu \nu}}{\log[r_H u_\infty]}\right\}.
\end{equation}

Using the equations (\ref{atEx},\ref{axEx}), and the asymptotic behaviour of the functions $S_{1,2}\sim \pm u/2$
and that the $D_i$ terms are subleading at the boundary, we have,
\begin{equation}
a_x(u\to \infty,k) = \frac{\omega E_x(u,k)}{\omega^2 - k^2} + a^0_x, \qquad a_t(u\to \infty,k) = \frac{-k
E_x(u,k)}{\omega^2 - k^2} + a^0_t,
\end{equation}
for some constants $a^0_\mu$.

Combining the above equations we have
\begin{equation}
k a_t(u \to \infty, k) + \omega a_x(u\to\infty,k) = E_x(u\to\infty,k) + \omega
a^0_x + k a^0_t.
\end{equation}
This implies that
\begin{equation}
\omega a^0_x + k a^0_t =0.
\end{equation}
As a result, at the boundary, the counter terms $A_\mu A^\mu$ evaluates to $E_x E_x/(\omega^2-k^2)$.

The contribution of the counter terms to the Green's functions are thus given by
\begin{eqnarray}
C_{\bar{e}e}&=& -|e_l|^2 (\log[u_\infty]+ \log r_H)-( \bar{e}_0 e_l+  e_0 \bar{e}_l) + |e_1|^2 , \nonumber \\
C_{\bar{e}t}&=& \frac{1}{2}\left((t_l \bar{e}_1 \log[u_\infty]+ t_1\bar{e}_1) - t_0 \bar{e}_l- t_l\bar{e}_1- \bar{e}_1(t_l-t_1)-\bar{e}_lt_0+\bar{e}_1 t_l \log[u_\infty]\right), \nonumber \\
C_{\bar{t}e}&=& \overline{C_{\bar{e}t}},   \nonumber \\
C_{\bar{t}t}&=& |t_l\log[u_\infty]+t_1|^2 -(|t_l|^2 (\log[u_\infty] + \log r_H)+ \bar{t}_lt_1 +\bar{t}_1t_l).
\end{eqnarray}
Note that the $\log r_H$ contribution is explicitly real, which we will drop from now (.i.e. setting $r_H=1$).

Combining with (\ref{bareGreen}), and restoring factors of $\omega^2-k^2$ we have
\begin{equation*}\label{Green}
\left(\begin{array}{cc}
G_{\bar{e}e} & G_{\bar{e}t}\\
G_{\bar{t}e} & G_{\bar{t}t}
\end{array}\right) =  \mathcal{N}\left(\begin{array}{cc}
\frac{2\bar{e}_le_0}{\omega^2-k^2}  & \frac{\bar{e}_lt_0}{\omega^2-k^2} + t_l\bar{e}_1 \\
\frac{e_l\bar{t}_0}{\omega^2-k^2} + \bar{t}_le_1 & 2\bar{t}_lt_1
\end{array}\right).
\end{equation*}
Putting back the correct normalisations as defined in (\ref{mixsole}, \ref{mixsolt}), we have finally

\begin{equation*}\label{Green2}
\left(\begin{array}{cc}
G_{\bar{e}e} & G_{\bar{e}t}\\
G_{\bar{t}e} & G_{\bar{t}t}
\end{array}\right) =  \mathcal{N}\left(\begin{array}{cc}
2\frac{e_0}{e_l(\omega^2-k^2)}  & \frac{t_0}{t_l(\omega^2-k^2)}+\frac{\bar{e}_1}{\bar{e}_l}\\
\frac{\bar{t}_0}{\bar{t}_l(\omega^2-k^2)}+\frac{e_1}{e_l} & 2\frac{t_1}{t_l}
\end{array}\right),
\end{equation*}
as advertised in (\ref{Greensmix}).


\bibliographystyle{JHEP}
\bibliography{d3d3bib}
\end{document}